\documentclass[12pt]{article}
\pdfoutput=1

\usepackage{draft} 
\usepackage{hyperref}
\usepackage{graphicx,color,subfig}
\usepackage{cite}
\usepackage{mciteplus}
\usepackage{skak}
\usepackage{empheq}
\usepackage{tikz}
\usepackage{booktabs}
\usepackage{siunitx}
\usepackage{bbm}
\usepackage{makecell}
\usepackage{float}

\usepackage{bm}
\usepackage{braket}

\DeclareFontFamily{OT1}{pzc}{}
\DeclareFontShape{OT1}{pzc}{m}{it}{<-> s * [1.10] pzcmi7t}{}
\DeclareMathAlphabet{\mathpzc}{OT1}{pzc}{m}{it}

\def\be#1\ee{\begin{align}#1\end{align}}

\makeatletter
\newcommand{\bdryno}{\mathpalette\bdry@no\relax}
\newcommand{\bdry@no}[2]{%
  \mspace{1mu}%
  \vbox{%
    \hbox{$\m@th#1\scriptstyle{\ast}$}
    \nointerlineskip
    \kern.25ex
    \hbox{$\m@th#1\scriptstyle{\ast}$}
    \kern-.06ex
  }%
  \mspace{1mu}%
}
\makeatother

\usetikzlibrary{snakes}
\usetikzlibrary{arrows}
\usetikzlibrary{decorations.pathmorphing}
\usetikzlibrary{ decorations.markings}
\tikzset{snake it/.style={decorate, decoration=snake}}
\usetikzlibrary{shapes.misc}
\tikzset{cross/.style={cross out, draw=black, minimum size=2*(#1-\pgflinewidth), inner sep=0pt, outer sep=0pt},
cross/.default={1pt}}

    \newcommand{\nssect}{\text{NS}}
\newcommand{\rsect}{\text{R}}

\newcommand{\upsns}{\Upsilon_{\nssect}}
\newcommand{\upsr}{\Upsilon_{\rsect}}

\newcommand{\zbar}{\overline{z}}
\newcommand{\cF}{\mathcal{F}}
\newcommand{\cT}{\mathcal{T}}
\newcommand{\cR}{\mathcal{R}}

\tikzset{
  pics/cylnnrb0/.style n args={2}{
    code = { %
        \filldraw[color=red,fill=black!30, thick] circle (0.35);
        \filldraw[color=blue, fill=white, thick]  circle (0.17);
        \node[color=black] at (0,0.51) {\scriptsize ${#1}$};
        \node[color=black] at (0,0) {\scriptsize ${#2}$};

    }
  }
}

\tikzset{
  pics/cylnnrb/.style n args={2}{
    code = { %
        \filldraw[color=red, densely dashed,fill=black!30, thick] circle (0.35);
        \filldraw[color=blue, densely dashed, fill=white, thick]  circle (0.17);
        \node[color=black] at (0,0.51) {\scriptsize ${#1}$};
        \node[color=black] at (0,0) {\scriptsize ${#2}$};

    }
  }
}
\tikzset{
  pics/cylnnrr/.style n args={2}{
    code = { %
        \filldraw[color=red, densely dashed,fill=black!30, thick] circle (0.35);
        \filldraw[color=red, densely dashed, fill=white, thick]  circle (0.17);
        \node[color=black] at (0,0.51) {\scriptsize ${#1}$};
        \node[color=black] at (0,0) {\scriptsize ${#2}$};

    }
  }
}
\tikzset{
  pics/cylnnbb/.style n args={2}{
    code = { %
        \filldraw[color=blue, densely dashed,fill=black!30, thick] circle (0.35);
        \filldraw[color=blue, densely dashed, fill=white, thick]  circle (0.17);
        \node[color=black] at (0,0.51) {\scriptsize ${#1}$};
        \node[color=black] at (0,0) {\scriptsize ${#2}$};

    }
  }
}

\tikzset{
  pics/disk1/.style n args={1}{
    code = { %
        \filldraw[color=black, fill=black!30, thick] circle (0.3);
        \node[color=black] at (0,0.5) {\scriptsize ${#1}$};
        \draw node[cross=3pt, very thick] {};
    }
  }
}

\begin{document}

\unitlength = .8mm

\begin{titlepage}

\begin{center}

\hfill \\
\hfill \\
\vskip 1cm

\title{The S-Matrix of 2D Type 0B String Theory
\\
Part 1: Perturbation Theory Revisited}

\author{Bruno Balthazar$^\diamondsuit$, Victor A. Rodriguez$^{\heartsuit}$, Xi Yin$^\spadesuit$}

\address{
$^\diamondsuit$Enrico Fermi Institute \& Kadanoff Center for Theoretical Physics,\\
University of Chicago, Chicago, IL 60637, USA \\
$^\heartsuit$Joseph Henry Laboratories, Princeton University, \\ Princeton, NJ 08544, USA \\
$^\spadesuit$Jefferson Physical Laboratory, Harvard University, \\
Cambridge, MA 02138 USA
}

\email{brunobalthazar@uchicago.edu, vrodriguez@princeton.edu, xiyin@fas.harvard.edu}

\end{center}

\abstract{We study the perturbative S-matrix of closed strings in the two-dimensional type 0B string theory from the worldsheet perspective, by directly integrating correlation functions of ${\cal N}=1$ Liouville theory. The latter is computed numerically using recurrence relations for super-Virasoro conformal blocks. We show that the tree level 3- and 4-point amplitudes are in agreement with the proposed dual matrix quantum mechanics. The non-perturbative aspects of the duality will be analyzed in a companion paper.}

\vfill

\end{titlepage}

\eject

\begingroup
\hypersetup{linkcolor=black}

\tableofcontents

\endgroup

%\pagebreak

\section{Introduction} 

The two-dimensional type 0B string theory, defined by a superconformal worldsheet theory that involves ${\cal N}=1$ Liouville theory and a diagonal GSO projection, was conjectured nearly two decades ago to be dual to a scaling limit of a gauged matrix quantum mechanics (MQM) \cite{Takayanagi:2003sm, Douglas:2003up}. While a number of evidences for this duality were presented at the level of the spectrum of string states and branes, much less has been established at the level of dynamics. Even at the leading nontrivial order in string perturbation theory, the only available results from the worldsheet have been obtained through ``resonance computation" \cite{DiFrancesco:1991daf} that involves a rather subtle analytic continuation in momenta (see \cite{Balthazar:2017mxh} for a justification in the context of the 2D bosonic ``$c=1$" string theory), rather than a first-principle calculation based on the actual correlators of the worldsheet CFT. The goal of the present paper is to fill this gap in the literature. In particular, we explicitly calculate the tree level $1\to 2$ and $1\to 3$ scattering amplitude of closed string states and show that they agree with results of the dual matrix model \cite{DeWolfe:2003qf}. 

The main technical ingredient of this paper, namely correlation functions in ${\cal N}=1$ Liouville theory, can be expressed as integrals of super-Virasoro conformal blocks against the structure constants previously obtained in \cite{Rashkov:1996np,Poghosian:1996dw,Fukuda:2002bv,Belavin:2007gz, Suchanek:2010kq}. The conformal blocks can be evaluated efficiently as an expansion in the plumbing parameter using the recurrence relations in central charge derived in \cite{Hadasz:2006qb,Belavin:2007gz}. In particular, we have numerically implemented the recurrence relation for the sphere four-point super-Virasoro blocks and verified the crossing invariance of four-point functions of generic primaries in ${\cal N}=1$ Liouville theory.

%The $1\to 3$ amplitude is particularly nontrivial as it involves four-point correlators of ${\cal N}=1$ Liouville theory, whose evaluation is made possible through a numerical implementation of the recursion relation for super-Virasoro conformal blocks.

The dual matrix quantum mechanics proposed by \cite{Takayanagi:2003sm, Douglas:2003up}, which we refer to as the type 0B MQM, is non-perturbatively defined and was thought to provide a non-perturbative completion of the 2D type 0B string theory. On the other hand, certain non-perturbative effects in string theory due to D-instantons are a priori  captured by worldsheets with boundaries \cite{Polchinski:1994fq, Green:1997tv}. This was recently made precise in the context of the bosonic $c=1$ string theory \cite{Balthazar:2019rnh, Balthazar:2019ypi, Sen:2019qqg}. In a companion paper \cite{paper:partii}, we will extend some of the results of \cite{Balthazar:2019rnh, Balthazar:2019ypi} to type 0B string theory, and provide evidence for the duality with the type 0B MQM at the non-perturbative level. 

In the next section, we will review the definition of the 2D type 0B string theory from the worldsheet perspective, as well as the dual matrix quantum mechanics. The conformal data of ${\cal N}=1$ Liouville theory are summarized in section \ref{sec:liouville}. The recursion formulae for ${\cal N}=1$ super-Virasoro conformal blocks involving NS sector states are given in section \ref{crecursions}.  The computation of tree level closed string amplitudes will be carried out in section \ref{perturbative}. Numerical consistency checks of the super-Virasoro blocks based on the crossing invariance of sphere four-point correlators of ${\cal N}=1$ Liouville theory are presented in Appendix \ref{crossing}. 
%Some numerical details in the evaluation of the four-point string amplitude are given in Appendices \ref{regularization} and \ref{numerics}.
Some details in the regularization of the four-point string amplitude are given in Appendix \ref{regularization}.

\section{An overview of type 0B string theory and its dual matrix quantum mechanics}
\label{sec:overview}

\subsection{The worldsheet description of two-dimensional type 0B string theory}
\label{sec:worldsheettheory}

The worldsheet description of the two-dimensional type 0B string theory, in the Neveu-Schwarz-Ramond (NSR) formalism, is based on a $(1,1)$ superconformal field theory (SCFT) that consists of a time-like free boson $X^0$ and its Majorana fermion partner $\psi^0, \widetilde\psi^0$, the $\cN=1$ Liouville theory of central charge $c={27\over 2}$, together with the $b,c,\beta,\gamma$ ghost system, subject to a diagonal GSO projection defined as follows.

The space of vertex operators in a two-dimensional $(1,1)$ SCFT a priori splits into R and NS sectors according to whether the OPE of the vertex operator with the holomorphic (left-moving) supercurrent $S$ and its anti-holomorphic (right-moving) counterpart has a two-fold branch cut or no branch cut. Each sector further admits a $\mathbb{Z}_2$-grading based on the fermion number symmetry under which the supercurrent is odd. We will refer to such sectors of even or odd fermion numbers as NS$\pm$ and R$\pm$, on the holomorphic and anti-holomorphic side independently.  Note that the left and right fermion numbers are exchanged under worldsheet parity symmetry. %For instance, the identity operator belongs to (NS+,NS+) sector, whereas $\psi^0$ and $\widetilde\psi^0$ belong to (NS-,NS+) and (NS+,NS-) sectors, respectively. 
The GSO projection that defines type 0B string theory retains the sectors
\ie
{\rm (NS+,NS+)} \oplus {\rm (NS-,NS-)} \oplus {\rm (R+,R+)} \oplus {\rm (R-,R-)}
\fe
as the space of admissible vertex operators.

The semiclassical action of ${\cal N}=1$ Liouville theory in flat space is given by
\ie
S_{\rm SL} = \int d^2z \left[ {1\over 2\pi} \left( \partial \phi \bar\partial \phi + \psi \bar\partial\psi + \widetilde\psi \partial \widetilde\psi \right) + 2i\mu b^2 \widetilde\psi \psi e^{b \phi} + 2\pi b^2 \mu^2 e^{2b \phi} \right],
\label{eq:SLsc}
\fe 
with $Q=b+b^{-1}$ and central charge $c={3\over 2} (1+2Q^2)$. This action includes a (supersymmetric) linear dilaton background which vanishes in flat space.  

The ${\cal N}=1$ Liouville theory will be reviewed in section \ref{sec:liouville}.  The operator spectrum in the (NS, NS) sector consists of super-Virasoro primaries $V_P$ labeled by the Liouville momentum $P\in \mathbb{R}_{\geq 0}$, of weight $h=\widetilde h = {1\over 2}({Q^2\over 4}+P^2)$. In the asymptotic region $\phi\to -\infty$ of the Liouville target space, the wave functional of $V_P$ can be identified with that of the free field operator
\ie\label{vpasymp}
& V_P \sim S_{\nssect}(P)^{-{1\over 2}} e^{({Q\over 2}+iP) \phi} + S_{\nssect}(P)^{1\over 2} e^{({Q\over 2}-iP) \phi}.
\fe
The explicit expression for the reflection $S_{\nssect}(P)$, given in \cite{Poghosian:1996dw}, will not be needed in this paper. % \left( \frac{\Gamma(iP)}{\Gamma(-iP)} \right)^2$ 

The (R, R) sector of ${\cal N}=1$ Liouville theory (viewed as a modular invariant bosonic CFT) consists of lowest weight states $V_P^{{\rm R},+}$ with respect to the superconformal algebra (SCA), of weight $h=\widetilde h = {1\over 16} + {1\over 2}({Q^2\over 4}+P^2)$, $P\geq 0$. In the asymptotic region $\phi\to -\infty$, the theory contains a weakly coupled fermion $(\psi^1,\widetilde\psi^1)$ in addition to the Liouville field $\phi$, and the wave functional of $V_P^{{\rm R}, +}$ in this limit can be identified with that of the free field operator
\ie\label{vpplusasymp}
& V_P^{{\rm R},+} \sim \sigma^1 ( S_{\rsect}(P)^{-{1\over 2}} e^{({Q\over 2}+iP) \phi} + S_{\rsect}(P)^{1\over 2} e^{({Q\over 2}-iP) \phi} ),
\fe
where $\sigma^1$ is the spin field in the $(\psi^1,\widetilde\psi^1)$-CFT. There is a closely related family of defect operators attached to a $\mathbb{Z}_2$ symmetry topological line, which we denote by $V_P^{{\rm R},-}$, with the $\phi\to -\infty$ asymptotics
\ie\label{vpminusasymp}
& V_P^{{\rm R},-} \sim \mu^1 ( S_{\rsect}(P)^{-{1\over 2}} e^{({Q\over 2}+iP) \phi} - S_{\rsect}(P)^{1\over 2} e^{({Q\over 2}-iP) \phi} ),
\fe
where $\mu^1$ is the disorder operator in the $(\psi^1,\widetilde\psi^1)$-CFT. Below we will specialize to the case $b=1$ and $Q=2$.
%The ${\cal N}=1$ Liouville theory is a modular invariant bosonic CFT if we view, say, $V_P^{{\rm R},+}$ as local operators whereas $V_P^{{\rm R},-}$ are viewed as defect operators attached to a $\mathbb{Z}_2$ topological defect line. 

%We note that in our definition of the Liouville vertex operators, such that they behave in the asymptotic region $\phi\to -\infty$ as in (\ref{vpasymp}, \ref{vpplusasymp}, \ref{vpminusasymp}), and hence their two-point functions are delta-function normalized (see equations (\ref{eq:normV}) and (\ref{eq:normVR}) in Section \ref{sec:liouville}) the reflection phases \cite{Poghosian:1996dw}
%\ie
%S_{\nssect}(P) = - \left( \frac{\Gamma(iP)}{\Gamma(-iP)} \right)^2, ~~~~~~
%S_{\rsect}(P) = \left( \frac{\Gamma(\frac{1}{2} + iP)}{\Gamma(\frac{1}{2} - iP)} \right)^2.
%\label{eq:legpoles}
%\fe
%have been absorbed into the NS and R sector vertex operators, respectively. It is in this convention that the three-point functions presented in this section are real-valued for real Liouville momenta, and in which there are no explicit ``leg-pole factors" in the dictionary between S-matrix elements of the full type 0B string theory and of the type 0B MQM.
%\fixme{end edit}

The full worldsheet SCFT admits a nilpotent BRST charge $Q_B$ such that the BRST transformation of the $b$ and $\B$ ghosts give the stress-energy tensor $T$ and the supercurrent $G$ respectively. The asymptotic 1-particle states of the string correspond to BRST cohomology classes subject to Siegel constraint. In the two-dimensional type 0B string theory, the only physical string states are those of two scalar particles in the spacetime, which we refer to as the ``tachyon" and the ``axion". They are represented by BRST-closed vertex operators (we work in $\A'=2$ convention)
\ie
\mathcal{T} ^{\pm}_{\omega} &= g_s c\widetilde c \, e^{-\phi-\widetilde{\phi}} e^{\pm i \omega X^0} V_{P=\omega}, \\
\cA^{\pm}_{\omega} &=  g_s \omega c\widetilde c  \, e^{-{\phi\over 2}-{\widetilde{\phi}\over 2}} e^{\pm i \omega X^0} A^{\pm}_{P=\omega},
\label{eq:vops}
\fe
where $\omega$ is the energy and the superscript $\pm$ on the LHS correspond to in- and out-states, respectively. 
%where we have bosonized the $\beta,\gamma$ ghosts via $\beta\simeq e^{-\phi}\partial\xi, \gamma\simeq e^{\phi}\eta$, as conventional, and 
$A_P^\pm$ is an (R, R) vertex operator of the matter SCFT, of the form
\ie
A^{\pm}_P = \frac{1}{\sqrt{2}}\left(  \sigma^0 V_P^{{\rm R}, +} \pm   \mu^0 V_P^{{\rm R},-} \right),
\fe
%where $R_P^{\pm\pm}$ %belongs to the (R$\pm$, R$\pm$) sector, and can be constructed as
%\ie{}
%& R_P^{++} = \sigma^0 V_P^{{\rm R}, +},~~~ R_P^{--} = \mu^0 V_P^{{\rm R},-},
%\fe
where $\sigma^0$ and $\mu^0$ are the spin field and disorder operator in the $(\psi^0,\widetilde\psi^0)$ free fermion CFT, and $V_P^{{\rm R},\pm}$ are the analogous operators in ${\cal N}=1$ Liouville theory introduced earlier. In particular, $A_P^{\pm}$ have weight $h=\widetilde h = {1\over 8} + {1\over 2}(1+P^2)$, and admit the following asymptotic description in the $\phi\to-\infty$ limit
\ie{}
& A_P^+\sim S^+\widetilde S^+  S(P)^{-{1\over 2}} e^{({Q\over 2}+iP) \phi} + S^- \widetilde S^-  S(P)^{{1\over 2}} e^{({Q\over 2}-iP) \phi},
\\
& A_P^-\sim S^-\widetilde S^-  S(P)^{-{1\over 2}} e^{({Q\over 2}+iP) \phi} + S^+ \widetilde S^+  S(P)^{{1\over 2}} e^{({Q\over 2}-iP) \phi},
\fe
where $S^\pm$ are two holomorphic spin fields in the $(\psi^0,\psi^1)$ system, and $\widetilde S^\pm$ the analogous anti-holomorphic spin fields.\footnote{Note the relation $S^\pm\widetilde S^\pm={1\over\sqrt{2}}(\sigma^0\sigma^1\pm \mu^0\mu^1)$ in the CFT of two free fermions.} 

Note that our convention for the phases of the NS and R vertex operators is such that the latter are delta-function normalized with real-valued three-point functions (see section \ref{sec:liouville}), and there will be no need to include ``leg-pole factors" in the relation between the S-matrix elements of the closed string states in the basis (\ref{eq:vops})
and of the dual MQM (as is in the consideration of \cite{Balthazar:2017mxh}).

%Here, $R^{++}_p$ refers to a primary in the R sector of the combined Liouville plus free boson system, has weight $h=\widetilde{h}=1/8 + 1/2(1+p^2)$, and has the same OPE coefficients as $R^{\pm}_p$ in the previous section. We can think of $R^{\pm\pm}_p\sim S^{\pm}\widetilde{S}^{\pm}V_p$, where $S^{\pm}$ are the two spin fields formed from the two fermions $\psi^0$ and $\psi^1$. 

%\fixme{$\bZ_2$ and scattering from left and right}

\subsection{The dual matrix quantum mechanics}

The proposed dual matrix model \cite{Takayanagi:2003sm, Douglas:2003up} is defined as a certain $N\to \infty$ limit of the $U(N)$ gauged matrix quantum mechanics with the Hamiltonian $H = {1\over 2}(P^2-X^2)$, where $X$ is a Hermitian $N\times N$ matrix and $P$ the conjugate canonical momentum matrix. It is a standard exercise to write the $U(N)$-invariant wave function as a function of the eigenvalues of $X$, on which the Hamiltonian acts as a similarity transformation of that of $N$ free fermions subject to the single-particle Hamiltonian $h = {1\over 2}(p^2-x^2)$. The scaling limit of interest is such that the ground state is described by fermions filling all energy eigenstates up to $E=-\mu$ ($\mu>0$), and one considers finite energy excitations thereof. In the semi-classical limit (large $\mu$), one may view the ground state as fermions occupying the region $h<-\mu$ in the phase space parameterized by $(x,p)$, which includes a region of $x<\sqrt{2\mu}$ and another region of $x>\sqrt{2\mu}$. We will refer to them as the LHS and the RHS of the fermi sea.

This matrix quantum mechanics is very similar to the so-called $c=1$ matrix model, which involves an identical Hamiltonian but a different choice of the ``closed string vacuum" state, where the fermi sea only occupies one side of the potential barrier. The latter is dual to $c=1$ string theory, defined by a bosonic worldsheet theory \cite{Klebanov:1991qa, Ginsparg:1993is, Jevicki:1993qn, Polchinski:1994mb}. In the present context, it is important that the fermi sea fills both sides of the potential.\footnote{Part of the original motivation in considering the two-sided fermi sea is to ensure non-perturbative stability. Note however that the ``one-sided" $c=1$ matrix model also admits a natural non-perturbative completion albeit without time-reversal symmetry \cite{Balthazar:2019rnh}.} The collective excitations of the fermi sea, which may be described as particle-hole pairs, or as ripples of the fermi surface in the semi-classical limit, are dual to closed strings in the two-dimensional type 0B string theory. The string coupling $g_s$ is proportional to $1/\mu$ (the precise relation will be given in section \ref{sec:resonance}).

\begin{figure}[h!]
\centering
\begin{tabular}{c c c}
\scalebox{0.75}{
\begin{tikzpicture}
\fill[white] (-5,-4) rectangle (5,-3/2);
\fill[black!30] (1.225,-4) rectangle (5,-3/2);
\draw[color=black, fill=black!30, line width=2pt] (1.225,-3/2) -- (5,-3/2);
\draw[black, line width=2pt, fill=white, domain=-2.005:2.005] plot (\x, {-\x*\x});
\draw[color=black, fill=black!30, line width=1pt] (0,0) -- (0,-3/2);
\draw[color=black, fill=black!30, line width=1pt] (-1/10,-3/2) -- (1/10,-3/2) node[right] {$-\mu$};
\end{tikzpicture}}
&&
\scalebox{0.75}{\begin{tikzpicture}
\fill[black!30] (-5,-4) rectangle (5,-3/2);
\draw[color=black, fill=black!30, line width=2pt] (1.225,-3/2) -- (5,-3/2);
\draw[color=black, fill=black!30, line width=2pt] (-1.225,-3/2) -- (-5,-3/2);
\draw[black, line width=2pt, fill=white, domain=-2.005:2.005] plot (\x, {-\x*\x});
\draw[color=black, fill=black!30, line width=1pt] (0,0) -- (0,-3/2);
\draw[color=black, fill=black!30, line width=1pt] (-1/10,-3/2) -- (1/10,-3/2) node[right] {$-\mu$};
\end{tikzpicture}} \\
$c=1$ MQM && type 0B MQM
\end{tabular}
\caption{Semi-classical picture of the vacuum state in $c=1$ MQM, in which we fill only one side of the potential barrier, and in type 0B MQM, in which \emph{both} sides of the potential are filled.
%\\Semi-classical picture of the vacuum in $c=1$ MQM and in type 0B MQM, in which we fill either one side or \emph{both} sides of the inverted quadratic potential, respectively.
}
\end{figure}
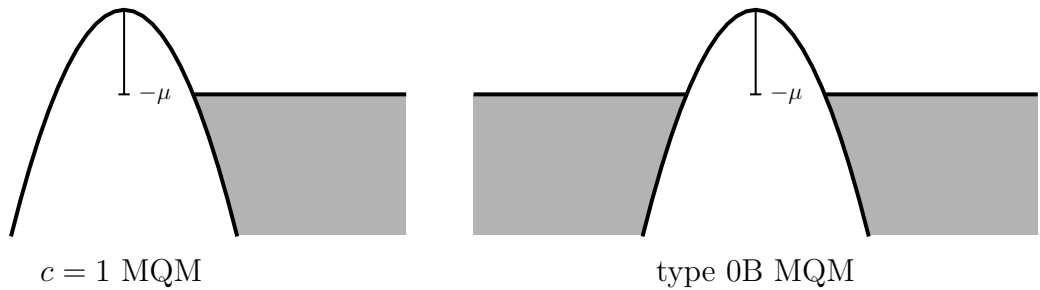

The separation of closed string states into modes associated with the LHS and the RHS of the fermi sea, while manifest in the proposed dual matrix quantum mechanics, is somewhat mysterious from the worldsheet perspective. The two sides of the fermi sea decouple at the level of perturbation theory, that is, the perturbative S-matrix is expected to factorize into the tensor product of those of the LHS modes and those of the RHS modes. We will see that the asymptotic states on the LHS of the fermi sea can be identified with the following linear combination of (\ref{eq:vops})
\ie
\cL^{\pm}_\omega = \cT^{\pm}_\omega - \cA^{\pm}_\omega, 
\label{eq:Lvertexop}
\fe
modulo the insertion of appropriate picture changing operators,
whereas the modes on the RHS of the fermi sea are identified with
\ie
\cR^{\pm}_\omega = \cT^{\pm}_\omega + \cA^{\pm}_\omega.
\label{eq:Rvertexop}
\fe
%In particular, the fermi sea on the left hand side of the inverted quadratic potential is perturbatively decoupled from the fermi sea on the right hand side. In other words, ${\cal A}^{\text{pert},(g)}_{1_{R/L}\to k_{L/R}}=0$, a fact that is manifest from a matrix quantum mechanics perspective, but not obvious from the string theory worldsheet perspective.

The exact S-matrix of the collective modes was obtained in \cite{Moore:1991zv} using the particle-hole representation. 
In particular, the S-matrix element of one collective mode being reflected into multiple collective modes on the same side of the fermi sea, 
\ie
S_{1_{R/L}\to k_{R/L}} (\omega, \omega_1,\cdots, \omega_k) = \delta\left(\omega - \sum_{i=1}^k \omega_i \right) \cA_{1_{R/L}\to k_{R/L}}(\omega_1,\ldots,\omega_k),
\fe
admits an asymptotic expansion in $\mu^{-1}$ of the form
\ie
\cA_{1_{R/L}\to k_{R/L}}(\omega_1,\ldots,\omega_k)&= \sum_{g=0}^{\infty}\frac{1}{\mu^{k-1+2g}}{\cal A}^{\text{pert},(g)}_{1_{R/L}\to k_{R/L}}(\omega_1,\ldots,\omega_k) + {\cal O}(e^{-2\pi \mu}).
\label{eq:mmRRinstexpansion}
\fe
This perturbative series is to be identified with the genus expansion of type 0B closed string amplitude, and is identical to (two copies of) that of $c=1$ string theory. Note however that in our convention there is a factor of 2 difference in the identification between the energy $\omega$ in the vertex operators (\ref{eq:vops}) and the energy $\omega_{\rm MM}$ of the collective mode in the matrix model, namely $\omega_{\rm MM}=2\omega$. For example, 
\ie
\cA^{\text{pert},(0)}_{1_R \to 2_R} &= 4 i \omega\omega_1\omega_2, \\
\cA^{\text{pert},(0)}_{1_R \to 3_R} &= 8 i \omega\omega_1\omega_2\omega_3 \left( 1+ 2 i\omega \right),
\label{eq:pertAmpsMM}
\fe
where we have included the factor of $1\over 2$ coming from the rescaling of the energy-conservation delta function. 
The main result of this paper will be to verify (\ref{eq:pertAmpsMM}) from the worldsheet computation.
The non-perturbative corrections of order ${\cal O}(e^{-2\pi\mu})$ on the RHS of (\ref{eq:mmRRinstexpansion}) are accounted for by D-instanton effects, and will be analyzed in \cite{paper:partii}.

\section{${\cal N}=1$ Liouville theory and superconformal blocks}

In this section we review the ${\cal N}=1$ Liouville CFT \cite{Rashkov:1996np,Poghosian:1996dw,Fukuda:2002bv,Belavin:2007gz, Suchanek:2010kq}, and present recurrence relations for ${\cal N}=1$ super-Virasoro conformal blocks for sphere 4-point function of NS sector primaries that allow for efficient numerical computation of the blocks. Here we treat ${\cal N}=1$ Liouville theory as a bosonic CFT, whose conformal data consist of a modular invariant spectrum of local operators defined with diagonal GSO projection. When viewed as a building block of the worldsheet SCFT of the non-critical type 0B string theory, a general string vertex operator may be built out of defect operators in ${\cal N}=1$ Liouville theory as well, as discussed in section \ref{sec:worldsheettheory}. Nonetheless, the relevant correlation functions of defect operators can be easily obtained from those of the local operators (in the bosonic CFT sense).

\subsection{The spectrum and structure constants}
\label{sec:liouville}

The spectrum of $\cN=1$ Liouville CFT of background charge $Q=b+b^{-1}$ and central charge $c={3\over 2}(1+2Q^2)$, in the (NS, NS) sector, consists of super-Virasoro primaries $V_P$ labeled by the Liouville momentum $P\in\bR_{\geq0}$ with comformal weight $h=\widetilde h={1\over 2}({Q^2\over 4}+P^2)$. We work in the convention in which the two-point functions of $V_P$ are normalized to
\ie
\langle V_{P_1}(z,\bar{z}) V_{P_2}(0) \rangle = \pi\frac{\delta(P_1-P_2)}{|z|^{2(h_1+h_2)}}.
\label{eq:normV}
\fe
The level-${1\over 2}$ superconformal descendants are denoted by
\ie
& \Lambda_P = G_{-{1\over 2}}V_P, ~~~~
 \widetilde{\Lambda}_P = \widetilde{G}_{-{1\over 2}}V_P, ~~~~ W_P = G_{-{1\over 2}}\widetilde{G}_{-{1\over 2}}V_P.
\fe
Our normalization convention for the supercurrent $G(z)$ is such that $\{G_r, G_s\} = 2L_{r+s} + {c\over 12}(4r^2-1)\delta_{r,-s}$.

As opposed to the non-supersymmetric case, there are two independent structure constants. They are associated with the three-point functions
\ie
 \langle V_{P_1}(z_1)V_{P_2}(z_2)V_{P_3}(z_3) \rangle &= \frac{C(P_1,P_2,P_3)}{|z_{12}|^{2(h_1+h_2-h_3)}|z_{13}|^{2(h_1+h_3-h_2)}|z_{23}|^{2(h_2+h_3-h_1)}}, \\
\langle W_{P_1}(z_1)V_{P_2}(z_2)V_{P_3}(z_3) \rangle &= \frac{ \widetilde{C}(P_1,P_2,P_3)}{|z_{12}|^{2(h_1+h_2-h_3+1/2)}|z_{13}|^{2(h_1+h_3-h_2+1/2)}|z_{23}|^{2(h_2+h_3-h_1-1/2)}},
\label{eq:3pts}
\fe
and have been bootstrapped in \cite{Rashkov:1996np,Poghosian:1996dw,Fukuda:2002bv,Belavin:2007gz}. Note that $\langle W_{P_1}(z,\bar{z}) W_{P_2}(0) \rangle = \pi (2h_1)^2 \delta(P_1-P_2)|z|^{-2(h_1+h_2+1)} $. The three-point functions of other  descendants are determined by (\ref{eq:3pts}) through superconformal Ward identities. For instance, 
$\langle W_{P_1}(0)W_{P_2}(1)W_{P_3}'(\infty) \rangle = (h_1+h_2+h_3-{1\over 2})^2\, \widetilde{C}(P_1,P_2,P_3)$.
%{|z_{12}|^{2(h_1+h_2-h_3+1/2)}|z_{13}|^{2(h_1+h_3-h_2+1/2)}|z_{23}|^{2(h_2+h_3-h_1+1/2)}}$.

We will now specialize to the case $b=1$ which is relevant for the worldsheet CFT of 2D type 0B string. The structure constants $C$ and $\widetilde C$ are given by
\ie
C(P_1,P_2,P_3) =\, & \frac{i}{2} \frac{1}{\upsns(1 + i(P_1 + P_2 + P_3))} \\
&\times \left[ \frac{\Gamma(1+iP_1)}{\Gamma(1-iP_1)}\frac{\upsns(2 iP_1)}{\upsns(1 + i(P_2 + P_3 - P_1))} \times (\text{2 permutations}) \right],\\
\widetilde{C}(P_1,P_2,P_3) = \, & i \frac{1}{\upsr(1 + i(P_1 + P_2 + P_3))} \\
&\times \left[ \frac{\Gamma(1+iP_1)}{\Gamma(1-iP_1)}\frac{\upsns(2 iP_1)}{\upsr(1 + i(P_2 + P_3 - P_1))} \times (\text{2 permutations}) \right].
\label{eq:superDOZZNS}
\fe
Here the functions $\upsns$ and $\upsr$ are defined as
\ie
\upsns(x) = \frac{\Gamma(x/2)}{\Gamma(1-x/2)}\left(\Upsilon_1\left(\frac{x}{2}\right)\right)^2,~~~~
\upsr(x) = \left( \Upsilon_1\left(\frac{x+1}{2} \right) \right)^2,
\fe
where $\Upsilon_1(x)$ denotes a special case of Barnes double Gamma function, in the same notation as in \cite{Balthazar:2018qdv}. 
%%% generic b \upsns and \upsr
%\ie
%& \Upsilon_{\nssect}(x) = \Upsilon_b\left( \frac{x}{2} \right)\Upsilon_b\left( \frac{x+Q}{2} \right),\\
%& \Upsilon_{\rsect}(x) = \Upsilon_b\left( \frac{x+b}{2} \right)\Upsilon_b\left( \frac{x+1/b}{2} \right),
%\fe
Note that (\ref{eq:superDOZZNS}) take real values for real Liouville momenta $P_i$, as is required by the unitarity of ${\cal N}=1$ Liouville theory at $b=1$. 
Another property of the NS structure constants (\ref{eq:superDOZZNS}) is that under analytic continuation of any of the momenta $P_i \to -P_i$, both $C(P_1,P_2,P_3)$ and $\widetilde{C}(P_1,P_2,P_3)$ flip sign.\footnote{The DOZZ structure constants for bosonic Liouville at $b=1$ also satisfiy this property \cite{Balthazar:2017mxh}.}

%%%% super-DOZZ for generic b %%%%
%{
%%structure constants for generic b 
%\color{gray}
%
%For generic $b$, the structure constants read
%\ie
%C(P_1,P_2,P_3) = & \left( \prod_{i=1}^3 (S_{\nssect}(P_i))^{-1/2} \right) \frac{\Upsilon'_b(0)}{\Upsilon_{\nssect}(Q/2 + i(P_1 + P_2 + P_3))} \\
%&\times \left[ \frac{\Upsilon_{\nssect}(Q + 2 iP_1)}{\Upsilon_{\nssect}(Q/2 + i(P_2 + P_3 - P_1))} \times (\text{2 permutations}) \right],\\
%& \vphantom{1}\\
%\tilde{C}(P_1,P_2,P_3) = & \left( \prod_{i=1}^3 (S_{\nssect}(P_i))^{-1/2} \right) \frac{2\Upsilon'_b(0)}{\Upsilon_{\rsect}(Q/2 + i(P_1 + P_2 + P_3))} \\
%&\times \left[ \frac{\Upsilon_{\nssect}(Q + 2 iP_1)}{\Upsilon_{\rsect}(Q/2 + i(P_2 + P_3 - P_1))} \times (\text{2 permutations}) \right],
%\label{eq:superDOZZgenericb}
%\fe
%where,
%\ie
%S_{\nssect}(P) = (b^{1 - b^2})^{2 i P/b} \frac{b^2 \Gamma(i b P) \Gamma(1 + i P/b)}{\Gamma(-i P/b) \Gamma(1 - i b P)}
%\fe
%One can check they are indeed real and reduce to (\ref{eq:superDOZZ}) as $b\to 1$.
%\\
%{[To do: Simplify as in (B.9) of 1702.00423 -- uniqueness of Liouville paper.]}
%% end of generic b structure constants
%}

The (R, R) sector Virasoro primaries $V_P^{{\rm R},+}$ and the defect operators $V_P^{{\rm R},-}$, with the asymptotic forms (\ref{vpplusasymp}) and (\ref{vpminusasymp}) respectively, have two-point functions normalized to
%spectrum consists of Virasoro primaries $V_P^{{\rm R},+}$, labeled by Liouville momentum $P\in\bR_{\geq 0}$, of weight $h'={1\over 16} + {1\over 2}({Q^2\over 4}+P^2)$ and the asymptotic form (\ref{vpplusasymp}). There is also a family of defect operators $V_P^{{\rm R},-}$ with the same weights, of the asymptotic form (\ref{vpminusasymp}). Both $V_P^{{\rm R},+}$ and $V_P^{{\rm R},-}$ will be used in constructing (R, R) vertex operators of string states. 
%Their two-point functions are normalized to 
\ie
\langle V_{P_1}^{{\rm R},\pm}(z,\bar{z}) V_{P_2}^{{\rm R},\pm}(0) \rangle = \pi\frac{\delta(P_1-P_2)}{|z|^{2(h_1'+h_2')}}.
\label{eq:normVR}
\fe
The two families of structure constants appearing in
\ie
\langle V_{P_3}(z_3)V_{P_2}^{{\rm R},+}(z_2)V_{P_1}^{{\rm R},+}(z_1) \rangle &= \frac{C^{+}(P_1,P_2;P_3) }{|z_{12}|^{2(h'_1+h'_2-h_3)}|z_{13}|^{2(h'_1+h_3-h'_2)}|z_{23}|^{2(h'_2+h_3-h'_1)}}, \\
\langle V_{P_3}(z_3)V_{P_2}^{{\rm R},-}(z_2)V_{P_1}^{{\rm R},-}(z_1) \rangle &= \frac{C^{-}(P_1,P_2;P_3) }{|z_{12}|^{2(h'_1+h'_2-h_3)}|z_{13}|^{2(h'_1+h_3-h'_2)}|z_{23}|^{2(h'_2+h_3-h'_1)}}, 
\fe
have been bootstrapped in \cite{Poghosian:1996dw, Suchanek:2010kq}, with the results given by
\ie
C^{+}(P_1,P_2;P_3) \equiv \frac{1}{2} \left( C^{\rm even}(P_1,P_2;P_3) + C^{\rm odd}(P_1,P_2;P_3) \right), \\
C^{-}(P_1,P_2;P_3) \equiv \frac{1}{2} \left( C^{\rm even}(P_1,P_2;P_3) - C^{\rm odd}(P_1,P_2;P_3) \right),
\fe
where 
\ie
C^{\rm even}(P_1,P_2;P_3) =\, & \frac{-i}{2}\frac{1}{\upsr(1+i(P_1+P_2+P_3))} \left[ \frac{\Gamma(1+iP_3)}{\Gamma(1-iP_3)}\frac{\upsns(2iP_3)}{\upsr(1 + i(P_1 + P_2 - P_3))} \right. \\
&~~~~~~~~~~~~~~\times \left. \frac{\Gamma(1/2+iP_1)}{\Gamma(1/2-iP_1)}\frac{\upsr(2iP_1)}{\upsns(1 + i(P_2 + P_3 - P_1))} \times (1\leftrightarrow 2)\right], \\
C^{\rm odd}(P_1,P_2;P_3) =\, & \frac{-i}{2}\frac{1}{\upsns(1+i(P_1+P_2+P_3))} \left[ \frac{\Gamma(1+iP_3)}{\Gamma(1-iP_3)}\frac{\upsns(2iP_3)}{\upsns(1 + i(P_1 + P_2 - P_3))} \right. \\
&~~~~~~~~~~~~~~\times \left. \frac{\Gamma(1/2+iP_1)}{\Gamma(1/2-iP_1)}\frac{\upsr(2iP_1)}{\upsr(1 + i(P_2 + P_3 - P_1))} \times (1\leftrightarrow 2)\right].
\label{eq:superDOZZR}
\fe
%in the case $b=1$. 
Once again, these structure constants are real-valued for $P_i\in \bR_{\geq 0}$. 

Note that upon analytic continuation of the momentum of the NS vertex operator $P_3 \to - P_3$, both $C^{\rm even}(P_1,P_2;P_3)$ and $C^{\rm odd}(P_1,P_2;P_3)$ flip sign. On the other hand, upon continuation of the momentum of either of the R vertex operators, $P_i \to -P_i$ with $i=1,2$, the R sector structure constants transform according to
\ie
C^{\rm even}(-P_1,P_2;P_3) = C^{\rm odd}(P_1,P_2;P_3), \\
C^{\rm odd}(-P_1,P_2;P_3) = C^{\rm even}(P_1,P_2;P_3).
\fe

%We note that with our normalization conventions (\ref{eq:normV}) and (\ref{eq:normVR}), in which the Liouville vertex operators behave in the asymptotic region $\phi\to -\infty$ of Liouville target space as in (\ref{vpplusasymp}), for example, the reflection phases 
%\ie
%S_{\nssect}(P) = - \left( \frac{\Gamma(iP)}{\Gamma(-iP)} \right)^2, ~~~~~~
%S_{\rsect}(P) = \left( \frac{\Gamma(\frac{1}{2} + iP)}{\Gamma(\frac{1}{2} - iP)} \right)^2.
%\label{eq:legpoles}
%\fe
%have been absorbed into the definition of the NS and R sector vertex operators, respectively. It is in this convention that the three-point functions presented in this section are real-valued for real Liouville momenta, and in which there are no explicit ``leg-pole factors" in the dictionary between S-matrix elements of type 0B string theory and of the type 0B MQM.

\subsection{Recursion relations for super-Virasoro conformal blocks}
\label{crecursions}

In this section we review the so-called ``$c$-recursion relations" of the $\cN=1$ super-Virasoro conformal blocks for the sphere 4-point function in the NS sector, derived in \cite{Hadasz:2006qb,Belavin:2007gz}\footnote{We note that there is a typo in the formulae presented in \cite{Hadasz:2006qb}, namely that the sign $\sigma^{rs}(\_h_3)$ appearing in the residue factor $R^m_{r,s}(h_4,\_h_3,\_h_2,h_1;h)$ in (\ref{eq:residuefactor}) was missed. With the sign $\sigma^{rs}(\_h_3)$ defined in the present paper, below equation (\ref{eq:residuefactor}), we have verified the crossing symmetry of sphere four-point correlation functions in $\cN=1$ Liouville theory in Appendix \ref{crossing}.} in close analogy with their bosonic counterpart \cite{Zamolodchikov:1985ie}, that will allow for efficient numerical evaluations.

We will parametrize the central charge of the superconformal algebra as $c={3\over 2}(1+2Q^2)$, with $Q=b+b^{-1}$, the holomorphic weights of external primaries as
\ie\label{hairel}
h_i = {1\over 2}({Q^2\over 4} - a_i^2), 
\fe
and the weight of the internal primary as $h = {1\over 2}({Q^2\over 4}+P^2)$. We further define 
\ie\label{bcrss}
(b_{r,s}(h))^2 &= -\frac{1}{r^2-1} \left( 4h+rs-1+\sqrt{16h^2+8(rs-1)h+(r-s)^2} \right), \\
c_{r,s}(h) &= \frac{15}{2} + 3b_{r,s}(h)^2 + 3b_{r,s}(h)^{-2}, \\
A_{r,s}(h) &= \frac{1}{2}\underset{\substack{(p,q)\neq (0,0),(r,s) \\ p+q \in 2\bZ}}{\prod_{p=1-r}^r \prod_{q=1-s}^s } \frac{\sqrt{2}}{pb_{r,s}+q b_{r,s}^{-1}}.
\fe
Our convention for the superconformal blocks will be such that the sphere correlation function of four scalar operators, either super-Virasoro primaries or their level-$({1\over 2},{1\over 2})$ descendants, can be decomposed as
\ie\label{eq:4pt}
&\langle \phi_4(z_4,\zbar_4) \phi_3(z_3,\zbar_3) \phi_2(z_2,\zbar_2) \phi_1(z_1,\zbar_1) \rangle 
\\
&= |z_{24}|^{-4h_2} |z_{14}|^{2(h_2-h_1+h_3-h_4)} |z_{34}|^{2(h_2+h_1-h_3-h_4)} |z_{13}|^{2(h_4-h_1-h_2-h_3)} 
 \sum_{h} C_{12h} C_{34h} |\cF(h_i;h|z)|^2, 
\fe
where $z=\frac{z_{12}z_{34}}{z_{13}z_{24}}$ is the cross ratio. 

%\fixme{even vs odd notation, $\widetilde{\_h}$ notation}
With all external operators in the NS sector, there are 8 independent superconformal blocks $\cF^{e/o}(h_4,h_3,h_2,h_1;h|z)$, $\cF^{e/o}(h_4,*h_3,h_2,h_1;h|z)$, $\cF^{e/o}(h_4,h_3,*h_2,h_1;h|z)$, and $\cF^{e/o}(h_4,*h_3,*h_2,h_1;h|z)$, where the superscript $e/o$ stands for even/odd, referring to the block receiving contribution from internal descendants of integer or half-integer levels, respectively.\footnote{A basis of super-Virasoro descendents of a primary state $|h\rangle$ (in the NS sector) is of the form $L_{-m_j}\cdots L_{-m_1}G_{-k_i}\cdots G_{-k_1}| h \rangle$ where $\{m_1,\ldots,m_j \}$ and $\{k_1,\ldots,k_i \}$ are two sets of positive integers and half-integers in increasing order, respectively.} The external operator may either be a superconformal primary of weight $h_i$, which we simply denote by $h_i$ in the argument of the superconformal block, or a level-${1\over 2}$ descendant of a weight $h_i$ primary, in which case we adopt the notation $*h_i$ in the argument of the block.\footnote{The notation $*h_i$ is a purely symbolic, in that a function with argument $*h_i$ is understood to be independent, a priori, from the function denoted by the same symbol but with argument $h_i$.}
%When appearing in a formula, the numerical value of $*h_i$ will be understood to be $h_i+{1\over 2}$. 
In an equation that applies to both types of external operators, we will denote both $h_i$ and $*h_i$ collectively as $\_ h_i$. In particular, all of the blocks admit series expansions in the cross ratio $z$ of the form\footnote{On the RHS, when $*h_i$ appears in the exponents its value is understood to be $h_i+{1\over 2}$, c.f. equations (\ref{eq:sCBs}) and (\ref{eq:ellipticblocks}).}
\ie\label{eq:sCBs}
\cF^{e}(h_4,\_h_3,\_h_2,h_1;h|z) &= z^{h-h_1-\_h_2} \left[ 1 + \sum_{m=\in\bZ_+} z^m F_{m}(h_4,\_h_3,\_h_2,h_1;h;c) \right], 
\\
\cF^{o}(h_4,\_h_3,\_h_2,h_1;h|z) &= \pm z^{h-h_1-\_h_2} \sum_{k\in\bZ_+ -{1\over 2}} z^k F_{k}(h_4,\_h_3,\_h_2,h_1;h;c) ,
\fe
where the overall sign on the RHS of the second line is $-$ in the case $(\_h_3,\_h_2)=(*h_3,*h_2)$, and $+$ otherwise.\footnote{Note that (\ref{eq:4pt}) does not specify the overall sign convention for the holomorphic superconformal blocks, which would be relevant for the correlators of level-$({1\over 2},0)$ descendants. An example is considered in Appendix \ref{crossing}.}

The coefficients of the $z$-expansion (\ref{eq:sCBs}) obey the following recursion relations
\ie\label{eq:NScrecursion}
&F_{0}(h_4,\_h_3,\_h_2,h_1;h;c)=1, \\
&F_{m}(h_4,\_h_3,\_h_2,h_1;h;c) 
= f_{m}(h_4,\_h_3,\_h_2,h_1;h) 
\\
& ~~~~~+ \underset{1<rs<2m,\,\, r+s=2\bZ}{\sum_{r=2,3,\ldots} \sum_{s=1,2,\ldots}} \frac{R^m_{r,s}(h_4,\_h_3,\_h_2,h_1;h)}{c-c_{r,s}(h)} F_{m-{rs\over 2}}(h_4,\_h_3,\_h_2,h_1;h+rs/2;c_{r,s}(h)).
\fe
The RHS involves a sum over poles of the superconformal block at $c=c_{r,s}(h)$ (defined in (\ref{bcrss})), whose residues are given by lower level Taylor coefficients of superconformal blocks at central charge $c_{r,s}(h)$ and shifted internal weight, multiplied by the factor
 \ie{}
&R^m_{r,s}(h_4,\_h_3,\_h_2,h_1;h) = 
\begin{cases}
      \sigma^{rs}(\_h_3) \left( -\frac{\partial c_{r,s}(h)}{\partial h} \right)A_{r,s}P_{r,s}(h_1,\_h_2)P_{r,s}(h_4,\_h_3) & \text{for } m\in\bZ_+, \\
      \sigma^{rs}(\_h_3) \left( -\frac{\partial c_{r,s}(h)}{\partial h} \right)A_{r,s}P_{r,s}(h_1,\widetilde{\_h_2})P_{r,s}(h_4,\widetilde{\_h_3}) & \text{for } m\in\bZ_+-\frac{1}{2},
\end{cases} 
\label{eq:residuefactor}
\fe
where $\sigma^{rs}(*h_3) = (-1)^{rs}$ and $\sigma^{rs}(h_3) = 1$, the tilde notation is $\widetilde{h} \equiv *h$ and $\widetilde{*h} \equiv h$, and the ``fusion polynomials" $P_{r,s}$ are defined as
\ie
P_{r,s}(h_1,h_2) &= \underset{(p+q)-(r+s)~\equiv~ 2\mod 4}{\prod_{ p \in \{1-r,r-1; 2 \} } \prod_{q\in \{1-s,s-1; 2\}  } } \frac{2a_1-2a_2-pb_{r,s}-q/b_{r,s}}{2\sqrt{2}}\,\frac{2a_1+2a_2+pb_{r,s}+q/b_{r,s}}{2\sqrt{2}}, \\
P_{r,s}(h_1,*h_2) &= \underset{(p+q)-(r+s)~\equiv~ 0\mod 4}{\prod_{ p \in \{1-r,r-1; 2 \} } \prod_{q\in \{1-s,s-1; 2\}  } } \frac{2a_1-2a_2-pb_{r,s}-q/b_{r,s}}{2\sqrt{2}}\,\frac{2a_1+2a_2+pb_{r,s}+q/b_{r,s}}{2\sqrt{2}},
\fe
where the notation $p \in \{1-r,r-1; 2 \}$ stands for $p$ ranging from $1-r$ to $r-1$ with step $2$, and $a_i$ are related to $h_i$ according to (\ref{hairel}).

The ``seed" block coefficient $f_m$ appearing on the RHS of (\ref{eq:NScrecursion}), which amounts to the $c\to\infty$ limit of $F_m$, is given by
\ie
f_{m}(h_4,h_3,h_2,h_1;h) &= 
\begin{cases}
      \frac{1}{m!} \frac{(h+h_3-h_4)_{m} (h+h_2-h_1)_{m}}{(2h)_{m}} & \text{for } m\in\bZ^{>0}, \\
      \frac{1}{(m-1/2)!} \frac{(h+h_3-h_4+1/2)_{m-1/2} (h+h_2-h_1+1/2)_{m-1/2}}{(2h)_{m+1/2}} & \text{for } m\in\bZ^{>0}-\frac{1}{2},\\
\end{cases} \\
f_{m}(h_4,h_3,*h_2,h_1;h) &= 
\begin{cases}
      \frac{1}{m!} \frac{(h+h_3-h_4)_{m} (h+h_2-h_1+1/2)_{m}}{(2h)_{m}} & \text{for } m\in\bZ^{>0}, \\
      \frac{1}{(m-1/2)!} \frac{(h+h_3-h_4+1/2)_{m-1/2} (h+h_2-h_1)_{m+1/2}}{(2h)_{m+1/2}} & \text{for } m\in\bZ^{>0}-\frac{1}{2},\\
\end{cases} \\
f_{m}(h_4,*h_3,h_2,h_1;h) &= 
\begin{cases}
      \frac{1}{m!} \frac{(h+h_3-h_4+1/2)_{m} (h+h_2-h_1)_{m}}{(2h)_{m}} & \text{for } m\in\bZ^{>0}, \\
      -\frac{1}{(m-1/2)!} \frac{(h+h_3-h_4)_{m+1/2} (h+h_2-h_1+1/2)_{m-1/2}}{(2h)_{m+1/2}} & \text{for } m\in\bZ^{>0}-\frac{1}{2},\\
\end{cases} \\
f_{m}(h_4,*h_3,*h_2,h_1;h) &= 
\begin{cases}
      \frac{1}{m!} \frac{(h+h_3-h_4+1/2)_{m} (h+h_2-h_1+1/2)_{m}}{(2h)_{m}} & \text{for } m\in\bZ^{>0}, \\
      -\frac{1}{(m-1/2)!} \frac{(h+h_3-h_4)_{m+1/2} (h+h_2-h_1)_{m+1/2}}{(2h)_{m+1/2}} & \text{for } m\in\bZ^{>0}-\frac{1}{2},\\
\end{cases}
\fe
where $(\cdot)_m$ stands for the Pochhammer symbol. 

%\fixme{change of variables to $q$}
In computing the sphere four-point string amplitude, we will be integrating the superconformal blocks, multiplied by the relevant structure constants, in the cross ratio $z$ over the complex plane. A priori, the $z$-expansion of the conformal blocks converges only within the unit disc $|z|<1$. However, one can extend the domain of analyticity to the entire complex $z$-plane, excluding $z=1,\infty$, \cite{1987TMP....73.1088Z, Maldacena:2015iua} by passing to the elliptic nome $q$ variable related to $z$ by
\ie\label{qnome}
q = \exp\left[ -\pi\frac{K(1-z)}{K(z)} \right], ~~~ \text{where } K(z) = {}_{2}F_{1}({1\over 2},{1\over 2},1|z).
\fe
More precisely, we can write the superconformal blocks (\ref{eq:sCBs}) as
\ie{}
\cF^{e/o}(h_4,\_h_3,\_h_2, h_1;h|z) &= (16q)^{h-Q^2/8} z^{Q^2/8-h_1-\_h_2} (1-z)^{Q^2/8-\_h_2 - \_h_3}  \theta_3(q)^{3/2Q^2-4(h_1 + \_h_2 + \_h_3 + h_4)} \\
& ~~~~\times H^{e/o}(h_4,\_h_3,\_h_2, h_1;h|q).
\label{eq:ellipticblocks}
\fe
The function $H^{e/o}(h_4,\_h_3,\_h_2, h_1;h|q)$, the so-called elliptic block, as a power series in $q$ converges in the unit $q$-disc, which covers the $z$-plane (and beyond) via (\ref{qnome}). Numerical evaluations will be performed by truncating the $q$-series expansion of the elliptic block.

\section{Worldsheet computation}
\label{perturbative}

We now perform the computation of tree level scattering amplitudes of 2D type 0B string from the worldsheet, based on the NSR formalism in which the integration over worldsheet gravitino, or Grassmann-odd moduli, is replaced by the insertion of picture-changing operators (PCOs) \cite{Friedan:1985ge, Polchinski:1998rr}. The holomorphic PCO takes the form
\ie
\chi = -\frac{1}{2} e^{\phi} \left( -i\psi^0 \partial X^0 + G^L \right) + \text{ghost terms},
\label{eq:pco}
\fe
where $G^L$ stands for the supercurrent of the ${\cal N}=1$ Liouville theory. For tree level (NS, NS) amplitudes, it suffices to arrange each pair of holomorphic and anti-holomorphic PCOs to coincide with a tachyon vertex operator, yielding the picture-raised version of the latter,\footnote{The overall factor of $i$ is due to the cocycle phase in converting $e^{-\phi}e^{- \widetilde\phi}$ to $e^{-\phi-\widetilde\phi}$ \cite{Polchinski:1998rq, Polchinski:1998rr}.}
\ie
\chi\widetilde{\chi}\cT^{\pm}_{\omega} &= {ig_s\over 4} \left[ \omega^2 \, \psi^0\widetilde{\psi}^0 e^{\pm i \omega X^0} V_{\omega} + e^{\pm i \omega X^0} W_{\omega} \pm \omega\, \widetilde{\psi}^0 e^{\pm i \omega X^0} \Lambda_{\omega} \mp \omega\, \psi^0 e^{\pm i \omega X^0} \widetilde{\Lambda}_{\omega} \right] + \cdots,
\label{eq:tachraised}
\fe
%where we have included a factor of $i$ in the first term so that every term in the square brackets has a positive two point function on the complex plane. 
where $\cdots$ represents additional terms involving ghosts that will not contribute to the string amplitudes considered in this section. In our convention, integrating out a pair of complex conjugate Grassmann odd moduli results in the insertion of $-i\chi\widetilde{\chi}$. 

\subsection{The $1\to 2$ amplitude}

The simplest nontrivial tree amplitude is that of a closed string decaying into two. Let us begin with the case in which all of the closed strings involved are ``RHS modes" of the fermi sea, represented by vertex operators of the form (\ref{eq:Rvertexop}). The tree amplitude for this ``$1_R \to 2_R$" process is computed by the three-punctured sphere diagram, 
\ie
S_{1_R\to 2_R}(\omega, \omega_1,\omega_2) ~=~& \langle c\widetilde{c}\cR^{+}_{\omega}(\infty) c\widetilde{c}\cR^{-}_{\omega_1}(1) c\widetilde{c}\cR^{-}_{\omega_2}(0) \rangle \\
=~& \langle c\widetilde{c}\left(-i\chi\widetilde{\chi}\right)\cT^{+}_{\omega}(\infty) c\widetilde{c}\cT^{-}_{\omega_1}(1) c\widetilde{c}\cT^{-}_{\omega_2}(0) \rangle + \langle c\widetilde{c}\cT^{+}_{\omega}(\infty) c\widetilde{c}\cA^{-}_{\omega_1}(1) c\widetilde{c}\cA^{-}_{\omega_2}(0) \rangle \\
&+ \langle c\widetilde{c}\cA^{+}_{\omega}(\infty) c\widetilde{c}\cT^{-}_{\omega_1}(1) c\widetilde{c}\cA^{-}_{\omega_2}(0) \rangle + \langle c\widetilde{c}\cA^{+}_{\omega}(\infty) c\widetilde{c}\cA^{-}_{\omega_1}(1) c\widetilde{c}\cT^{-}_{\omega_2}(0) \rangle 
\fe
where in the first term after the second equality we have inserted a pair of PCOs coincident with the first tachyon vertex operator, so that the total picture number on the sphere is $(-2,-2)$; the remaining three terms do not come with PCO insertions. The amplitude evaluates to a linear combination of the structure constants of $\cN=1$ Liouville theory introduced in Section \ref{sec:liouville}, 
\ie
S_{1_R\to 2_R}(\omega, \omega_1,\omega_2) = i\delta(\omega-\omega_1-\omega_2)&\, g_s^3 C_{S^2} \left[ \frac{1}{4} \widetilde{C}(\omega,\omega_1,\omega_2) + \frac{1}{2} \omega_1\omega_2 C^{\rm even}(\omega_1,\omega_2;\omega) \vphantom{\frac{1}{1}}\right. \\ 
&+ \left. \frac{1}{2} \omega\omega_2 C^{\rm odd}(\omega,\omega_2;\omega_1) + \frac{1}{2} \omega\omega_1 C^{\rm odd}(\omega,\omega_1;\omega_2) \vphantom{\frac{1}{1}}\right].
\fe
The formulae for the structure constants  (\ref{eq:superDOZZNS}) and (\ref{eq:superDOZZR}) simplify dramatically for the energy-conserving kinematic configuration $\omega=\omega_1+\omega_2$, giving the result (after factoring out the delta function in energy)
\ie
\cA_{1_R\to 2_R}(\omega_1,\omega_2) &= i g_s^3 C_{S^2}\, \omega \omega_1 \omega_2.
\fe
%{\fixme{note}\color{gray} Perturbative decoupling of LHS and RHS of MM inverted quadratic potential implies $\myc_2 \myc_3^2 = 1/2$. For example, we can set $\myc_2=1$ and $\myc_3=1/\sqrt{2}$, so that $A_p^{\pm}=1/2(R^{++}_p \pm R^{--}_p)$.}
Indeed, we recover the anticipated matrix model result (\ref{eq:pertAmpsMM}) upon the identification
\ie
g_s^3 C_{S^2} = {4\over \mu}.
\label{eq:const1to2}
\fe

On the other hand, one can verify that any tree-level three-point amplitude that involves both a RHS mode $\cR_{\omega}$ and a LHS mode $\cL_{\omega'}$ vanishes identically, as anticipated from the perturbative decoupling between the LHS and RHS modes that is manifest in the dual matrix model.
For instance, a putative string tree-level amplitude of the decay of a LHS mode into two RHS modes is computed by the following three-punctured sphere diagram,
\ie
S_{1_L\to 2_R}(\omega, \omega_1,\omega_2) ~=~& \langle c\widetilde{c}\cL^{+}_{\omega}(\infty) c\widetilde{c}\cR^{-}_{\omega_1}(1) c\widetilde{c}\cR^{-}_{\omega_2}(0) \rangle \\
=~& \langle c\widetilde{c}\left(-i\chi\widetilde{\chi}\right)\cT^{+}_{\omega}(\infty) c\widetilde{c}\cT^{-}_{\omega_1}(1) c\widetilde{c}\cT^{-}_{\omega_2}(0) \rangle + \langle c\widetilde{c}\cT^{+}_{\omega}(\infty) c\widetilde{c}\cA^{-}_{\omega_1}(1) c\widetilde{c}\cA^{-}_{\omega_2}(0) \rangle \\
&- \langle c\widetilde{c}\cA^{+}_{\omega}(\infty) c\widetilde{c}\cT^{-}_{\omega_1}(1) c\widetilde{c}\cA^{-}_{\omega_2}(0) \rangle - \langle c\widetilde{c}\cA^{+}_{\omega}(\infty) c\widetilde{c}\cA^{-}_{\omega_1}(1) c\widetilde{c}\cT^{-}_{\omega_2}(0) \rangle \\
=~& i\delta(\omega-\omega_1-\omega_2)\, g_s^3 C_{S^2} \left( \frac{1}{4} + \frac{1}{4} - \frac{1}{4} - \frac{1}{4} \right) \omega \omega_1 \omega_2 \\
=~& 0.
\fe
Note that the perturbative decoupling of LHS and RHS modes demands that various $1\to k$ amplitudes in the basis of tachyons and axions are equal in magnitude, %In the example considered above, the four different contributions simplified to $\frac{1}{4}\omega\omega_1\omega_2$. 
implying a rather intricate relationship between the NS and R sectors of the type 0B string theory. It appears that this relationship cannot be understood as a symmetry of the worldsheet CFT, as the relations among the correlators in question only hold after imposing energy conservation for the external string states, or more generally after integration over the moduli space of the worldsheet diagram. %Hence, this relationship between the NS and the R sector is of a stringy nature.
Nonetheless, we shall assume this property in the next section to simplify our calculation of the $1\to 3$ string amplitude.

%\fixme{crossing symmetry of $\cA_{1\to 2}$}
The $1\to 2$ amplitude admits a crossing symmetry upon analytic continuation in the energies, of the form
\ie
\cA_{2_R\to 1_{R}}(\{ \omega_1 , \omega_2 \},\{\omega_3 \}) = - \cA_{1_R\to 2_{R}}(\{ \omega_1 \},\{ -\omega_2 , \omega_3 \}).
\label{eq:1to2crossing}
\fe
This can be understood as a consequence of the property of the Liouville structure constants under continuation in the Liouville momenta $P_i \to -P_i$, as discussed in Section \ref{sec:liouville}. 
The appearance of the overall minus sign on the RHS of (\ref{eq:1to2crossing}), perhaps slightly unconventional, was also seen in the crossing relation of perturbative amplitudes in $c=1$ string theory \cite{Balthazar:2017mxh}.

\subsection{The $1\to 3$ amplitude}

Next, we consider the tree-level decay of a closed string into three closed strings, whose worldsheet computation is far more involved.
%A much more nontrivial test of the duality between the type 0B MQM and the type 0B string theory is the calculation of the tree-level decay of a closed string into three closed strings, say of RHS modes. 
%This calculation is made possible by the recurrence relations of super-Virasoro conformal blocks, reviewed in Appendix \ref{crecursions}, which allow for the efficient computation to high precision of the distinct blocks appearing in the sphere correlators.
We will focus on the RHS modes. The amplitude is computed by the four-punctured sphere diagram, integrated over its moduli space parameterized by the cross ratio $z$,
\ie
S_{1_R\to 3_R}(\omega, \omega_1,\omega_2,\omega_3) ~=~& \int_{\bC}d^2z \langle c\widetilde{c}\cR^{+}_{\omega}(\infty) c\widetilde{c}\cR^{-}_{\omega_3}(1) \cR^{-}_{\omega_1}(z,\zbar) c\widetilde{c}\cR^{-}_{\omega_1}(0) \rangle.
\label{eq:1to3amp}
\fe
It is understood that the RHS is expanded as a linear combination of correlators of tachyon and axion vertex operators, via  (\ref{eq:Rvertexop}) with the appropriate PCO insertions. There are 8 such terms that are nonzero, involving an even number of axion vertex operators, that must be equal to one another as required by the perturbative decoupling of the RHS modes from the LHS modes. %More precisely, the eight nonzero terms (those with an even number of axion vertex operator insertions) out of the total sixteen terms in the expansion must be equal to each other. 
We will proceed by assuming this, and will only explicitly evaluate the sphere diagram with four tachyon vertex operators. The latter computation involves only NS sector states and can be performed using the recurrence relations for the super-Virasoro conformal blocks presented in section \ref{crecursions}. %---and multiply the answer by a factor of 8 to obtain the complete amplitude of the decay of a RHS mode into three RHS modes. 
This gives
\ie
&S_{1_R\to 3_R}(\omega, \omega_1,\omega_2,\omega_3) = - 8 \int_{\bC}d^2z \langle c\widetilde{c}\cT^{+}_{\omega}(\infty) c\widetilde{c}\left(-i\chi\widetilde{\chi}\right)\cT^{-}_{\omega_3}(1) \left(-i\chi\widetilde{\chi}\right)\cT^{-}_{\omega_1}(z,\zbar) c\widetilde{c}\cT^{-}_{\omega_1}(0) \rangle \\
&= - 8i\delta\left(\omega-\sum_{i=1}^3 \omega_i\right)\, g_s^4 C_{S^2}  \\ 
&~~~\times \int_\bC d^2z |z|^{-2\omega_1\omega_2}|1-z|^{-2\omega_2\omega_3} {1\over 2^4} \bigg\langle V_{\omega}(\infty) \left( \omega_3^2 \, \psi^0\widetilde{\psi}^0 V_{\omega_3} + W_{\omega_3} - \omega_3\, \widetilde{\psi}^0 \Lambda_{\omega_3} + \omega_3\, \psi^0 \widetilde{\Lambda}_{\omega_3} \right)(1) \\
&~~~~~~~~~\left( \omega_2^2 \, \psi^0\widetilde{\psi}^0 V_{\omega_2} + W_{\omega_2} - \omega_2\, \widetilde{\psi}^0 \Lambda_{\omega_2} + \omega_2\, \psi^0 \widetilde{\Lambda}_{\omega_2} \right)(z,\zbar) V_{\omega_1}(0) \bigg\rangle \\
&= {i\over 2} \delta\left(\omega-\sum_{i=1}^3 \omega_i\right)\, g_s^4 C_{S^2} \bigg\{ \omega_2^2\omega_3^2\int_\bC d^2z |z|^{-2\omega_1\omega_2}|1-z|^{-2\omega_2\omega_3-2}\langle V_\omega(\infty) V_{\omega_3}(1) V_{\omega_2}(z,\zbar) V_{\omega_1}(0) \rangle   \\
&~~~~ - \int_\bC d^2z |z|^{-2\omega_1\omega_2}|1-z|^{-2\omega_2\omega_3}\langle V_\omega(\infty) W_{\omega_3}(1) W_{\omega_2}(z,\zbar) V_{\omega_1}(0) \rangle \\
&~~~~  - \omega_2\omega_3 \int_\bC d^2z |z|^{-2\omega_1\omega_2}|1-z|^{-2\omega_2\omega_3}\left[\frac{1}{1-\zbar}\langle V_\omega(\infty) \Lambda_{\omega_3}(1) \Lambda_{\omega_2}(z,\zbar) V_{\omega_1}(0) \rangle \right. \\
& ~~~~~~~~~~~~~~~~~~~~~~~~~~~~~~~~~~~~~~~~~~~~~~~~~~ +\left. \frac{1}{1-z}\langle V_\omega(\infty) \widetilde{\Lambda}_{\omega_3}(1) \widetilde{\Lambda}_{\omega_2}(z,\zbar) V_{\omega_1}(0) \rangle \right] \bigg\}.
\label{eq:1to3ampnaive}
\fe
The overall factor of 8 is due to summing over additional contributions involving axions. The overall minus sign is fixed by consideration of tree-level unitarity, namely the factorization property explained in \cite{Balthazar:2017mxh}. Alternatively, it can be determined by a direct comparison with the anticipated matrix model result (\ref{eq:pertAmpsMM}) which is manifestly unitary. %This results in the overall minus sign in the first line of (\ref{eq:1to3ampnaive}).

The integral on the RHS of (\ref{eq:1to3ampnaive}) diverges near the boundary of the moduli space where vertex operators collide, as is typically encountered in string perturbation theory. Conventionally, such divergences are dealt with by defining the string amplitude via analytic continuation in the external momenta from a domain where the moduli integral is finite. For the purpose of numerical evaluation, we will instead employ the regularization method introduced in \cite{Balthazar:2017mxh}, explicitly subtracting off power divergences while maintaining the analyticity in momenta. 
The detailed form of the divergences and the subtraction of counter terms are discussed in Appendix \ref{regularization}. 
The fully regularized, and manifestly finite, expression for the $1\to 3$ amplitude is given by (\ref{eq:1to3ampreg}), which we will evaluate numerically in section \ref{numerical1to3}.

\subsubsection{A resonance computation}
\label{sec:resonance}

A special case of the $1\to 3$ amplitude, where we analytically continue the energy of the incoming closed string to $\omega= i$, can be evaluated analytically analogously to the resonance computation discussed in \cite{Balthazar:2017mxh} in the context of $c=1$ string theory, as follows.

The relevant 4-point correlators of ${\cal N}=1$ Liouville theory admits a superconformal block decomposition that takes the form of an integral over the internal Liouville momentum $P$ as described in (\ref{eq:LiouvVVVV}--\ref{eq:LiouvVLLV}). The structure constant $C(\omega,\omega_3,P)$ or $\widetilde{C}(\omega,\omega_3,P)$, appearing in the coefficient of the integrand, has a double zero at $\omega=i$. Naively, one might conclude that the amplitude vanishes in this limit. On the other hand, as we analytically continue $\omega\to i$, we must deform the integration contour in $P$ accordingly to avoid singularities of the integrand (or account for the residue contribution when a pole crosses). In fact, a pair of poles of the structure constant $C(\omega,\omega_3,P)$, at $P=\pm i\mp \omega+\omega_3 \to \omega_3$, pinch the $P$ integration contour. 

Furthermore, the structure constant $C(\omega_2,\omega_1,P)$, also appearing in the coefficient of the integrand, has a pole at  $P= i - \omega_1 - \omega_2 \to \omega_3$.
The net effect is such that the pinch singularity and the extra pole offset the double zero at $\omega=i$ and yields a nonzero result, while the contribution from the rest of the contour away from the pinch singularity vanishes. 
On the other hand, no poles of the structure constant $\widetilde{C}(\omega,\omega_3,P)$ pinch the $P$-contour, and thus the terms in (\ref{eq:LiouvVVVV}--\ref{eq:LiouvVLLV}) involving $\widetilde C$ do not contribute at $\omega=i$.
%Therefore, as we continue to $\omega\to i$ only the contributions proportional to $C(\omega,\omega_3,P)C(\omega_2,\omega_1,P)$ in the conformal block expansion of the Liouville correlators (\ref{eq:LiouvVVVV}--\ref{eq:LiouvVLLV}) are non-vanishing.

The super-Virasoro conformal blocks in question, in the limit $\omega\to i$, reduce to those of a (super-)linear dilaton theory (with the same background charge as the ${\cal N}=1$ Liouville theory), with the simple expressions
\ie
\cF^{e}(h_1,h_2,h_3,h_{\text{in}}\to 0;h_P \to h_3;z) &\longrightarrow z^{-\alpha_1\alpha_2} (1-z)^{-\alpha_2\alpha_3}, \\
\cF^{o}(h_1,*h_2,*h_3,h_{\text{in}}\to 0;h_P \to h_3;z) &\longrightarrow  \alpha_2\alpha_3 \, z^{-\alpha_1\alpha_2} (1-z)^{-\alpha_2\alpha_3 - 1},
\fe
where $\alpha_i = 1 + i \omega_i$. It follows that the resonance $1_R \to 3_R$ amplitude at $\omega=i$ evaluates to
\ie{}
&\cA_{1_R\to 3_R}(\omega_1,\omega_2,\omega_3)\bigg|_{\omega_1+\omega_2+\omega_3=i} \\
& = {i\over 2} g_s^4 C_{S^2} \left( \omega_2^2\omega_3^2 + \alpha_2^2\alpha_3^2 + 2\omega_2\omega_3\alpha_2\alpha_3 \vphantom{\sum}\right)\\
& ~~~~ \times\underset{P\to -i+(\omega + \omega_3)}{(-2i)\text{Res}}\left[ C(\omega_3,\omega,P)C(\omega_1,\omega_2,P) \vphantom{\sum}\right] \int_\bC d^2z |z|^{-2\omega_1\omega_2-2\alpha_1\alpha_2}|1-z|^{-2\omega_2\omega_3-2\alpha_2\alpha_3-2}\\
&= {i\over 2} g_s^4 C_{S^2} \left( \omega_2^2\omega_3^2 + \alpha_2^2\alpha_3^2 + 2\omega_2\omega_3\alpha_2\alpha_3 \vphantom{\sum}\right) i\pi \frac{1}{\omega_1^2}\omega_1\omega_2\omega_3\\
& =  g_s^4 C_{S^2} \frac{\pi}{2}\omega_1\omega_2\omega_3.
\fe
Comparing with (\ref{eq:pertAmpsMM}), we reproduce the anticipated MQM result provided the identification
\ie
{8\over \mu^2} = g_s^4 C_{S^2} \frac{\pi}{2}.
\label{eq:const1to3}
\fe
It follows from (\ref{eq:const1to2}) and (\ref{eq:const1to3}) that the string coupling $g_s$, the normalization constant $C_{S^2}$ associated with the sphere topology, and the MQM parameter $\mu$ are identified via
\ie
g_s = \frac{4}{\pi\,\mu}, ~~~~~ C_{S^2} = \frac{\pi}{g_s^2}.
\label{eq:gsCS2}
\fe

\subsubsection{Numerical computation at complex energies}
\label{numerical1to3}

The numerical evaluation of the regularized $1\to 3$ amplitude (\ref{eq:1to3ampreg}) follows closely the strategy employed in \cite{Balthazar:2017mxh}. First of all, we evaluate the super-Virasoro blocks by truncating their expansions in the elliptic nome $q$ to a certain finite order. To reduce numerical errors, we would like to restrict the moduli integration to a reasonably small domain near the origin of the $q$-disc, or of the $z$-plane. This can be achieved using the crossing symmetry of the Liouville four-point functions (see Appendix \ref{crossing}, as well as Appendix C.2 of \cite{Chang:2014jta}). %In contrast to the consideration of \cite{Balthazar:2017mxh}, however, here we have fewer useful crossing relations available.
%crossing symmetry was used to reduce the moduli integration to the domain $\{|1-z|<1, 0<\text{Re}(z) <1/2 \}$. In this paper, 
In particular, (\ref{eq:Scrossing}) relates the relevant four-point functions at cross ratio $z$ to those at $1/z$, thereby reducing the moduli integration domain to the unit disc $D=\{|z| \leq 1 \}$.

Next, we split $D$ into three regions: (1) $D^{(0)}_\epsilon = \{|z| < \epsilon \}$, (2) $D_{\text{int}} = \{ \epsilon \leq |z| \leq 1, |1-z| \geq \epsilon \}$, and (3) $D^{(1)}_\epsilon = \{|z| \leq 1, |1-z| < \epsilon \}$, for a sufficiently small positive $\epsilon$.%, as shown in Figure \ref{fig:modspace}.
%Results shown in Figure \ref{fig:1to3ampWSvsMM} were computed for $\epsilon=1/100$.
The integration $D_{\text{int}}$ will be evaluated numerically using the $c$-recursion superconformal blocks of Section \ref{crecursions}. 
In region $D^{(0)}_\epsilon$, we truncate the superconformal blocks to low orders in $z$ and perform the integration analytically, while taking into account the counter terms.
In region $D^{(1)}_\epsilon$, we use the cross channel superconformal block decomposition such as (\ref{eq:TcrossingVWWV}) and (\ref{eq:TcrossingVLLV}). The relevant blocks in this case are of the type $\cF^{e/o}(h_4,h_3,*h_2,*h_1;h_P;1-z)$. While they may be determined from the blocks presented in section \ref{crecursions} through Ward identities, for our purpose it suffices to directly compute these blocks to low orders in $(1-z)$, as in (\ref{eq:coeffcpdef}) and (\ref{eq:coeffcp}).

\begin{figure}[h!]
\centering
\begin{tabular}{c c c}
\includegraphics[width=0.4\textwidth]{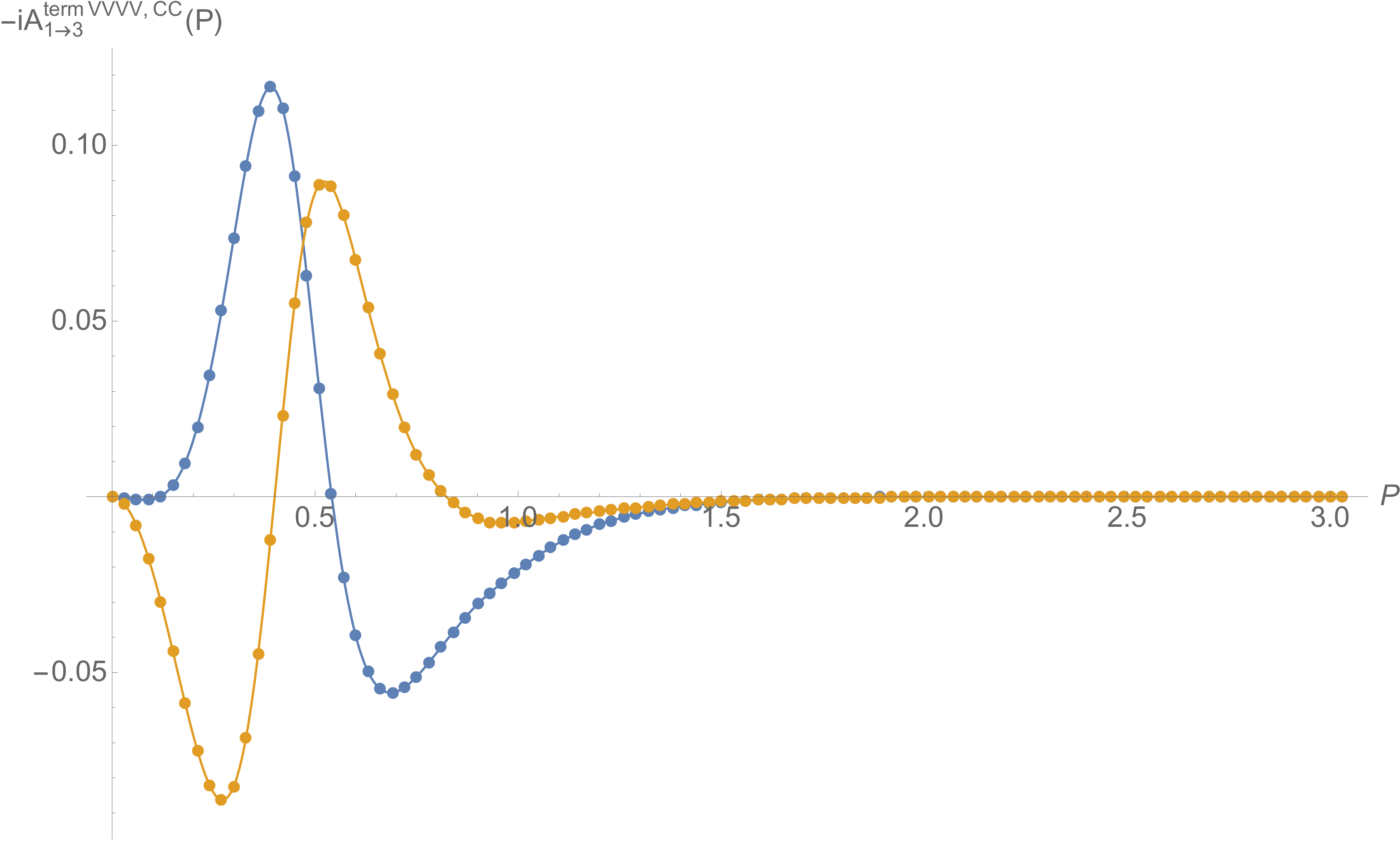} &~~~& \includegraphics[width=0.4\textwidth]{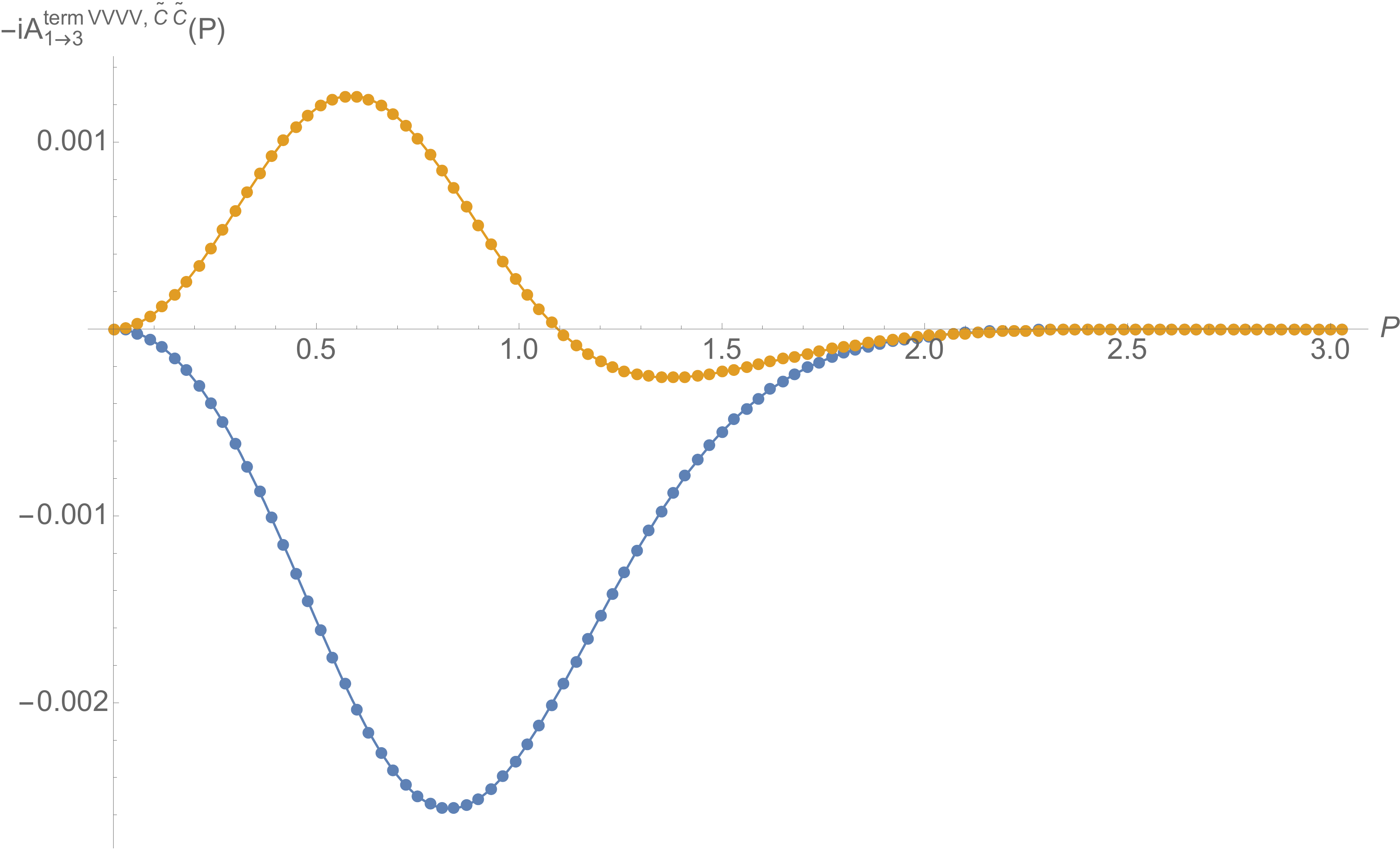} \\
\includegraphics[width=0.4\textwidth]{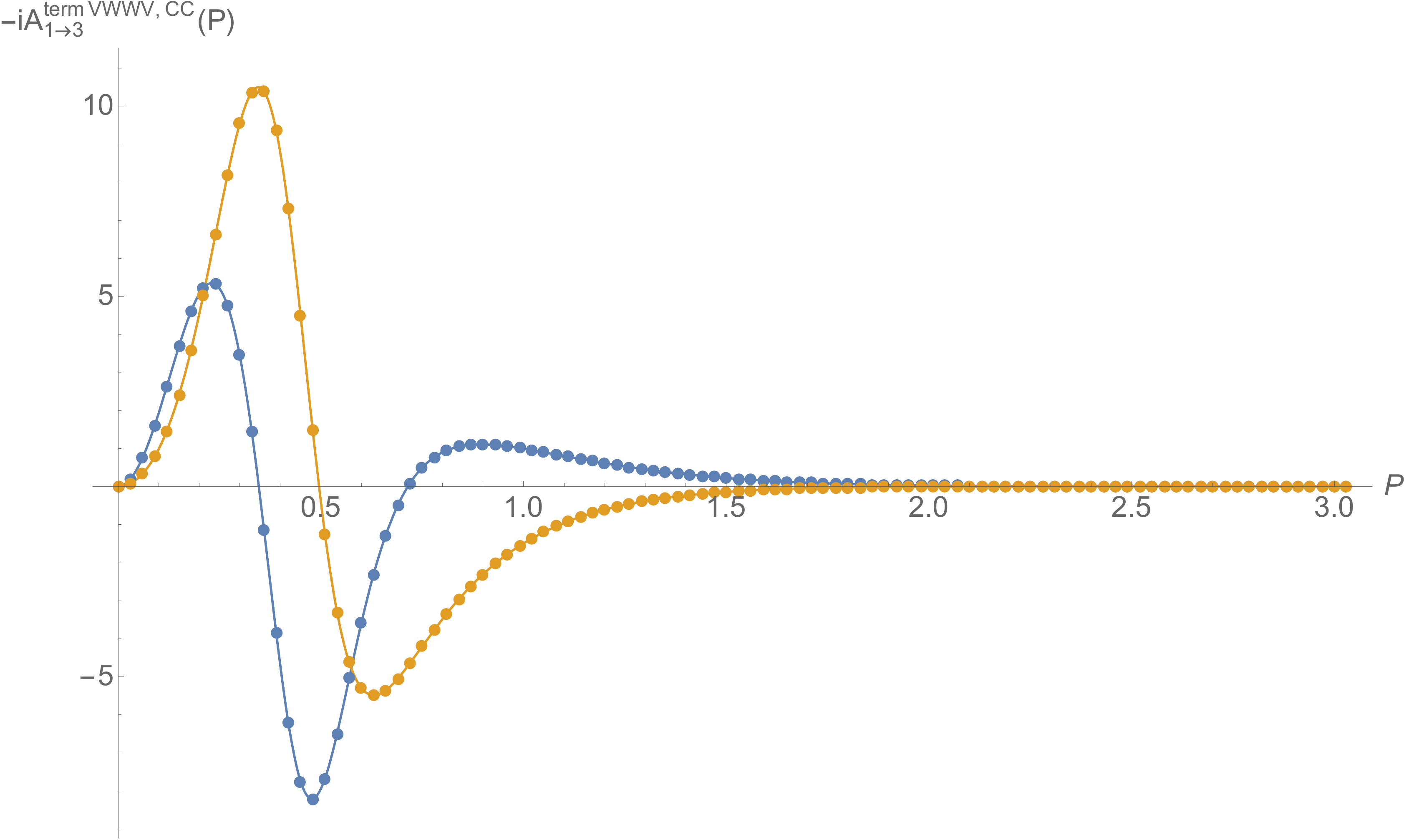} &~~~& \includegraphics[width=0.4\textwidth]{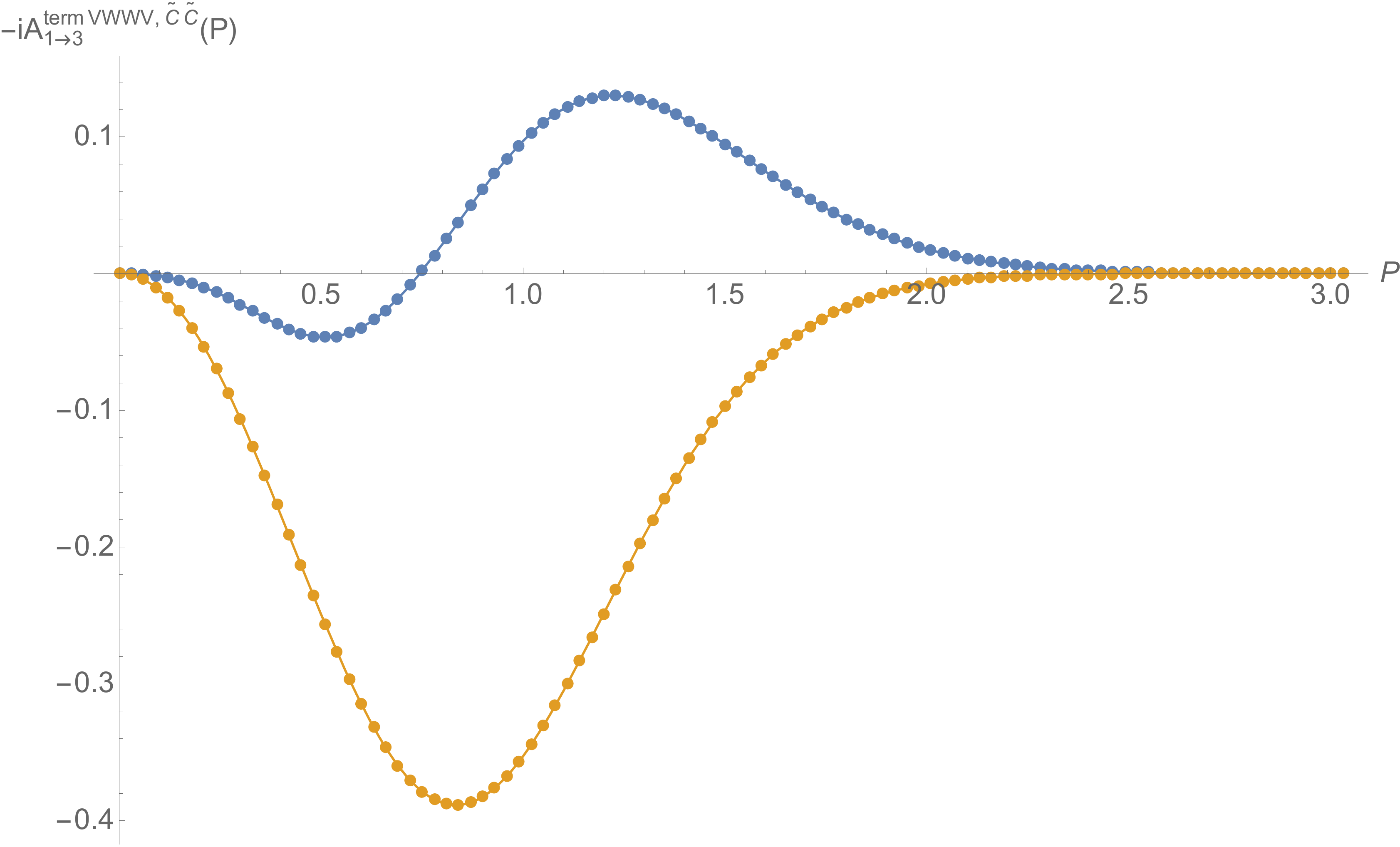} \\
\includegraphics[width=0.4\textwidth]{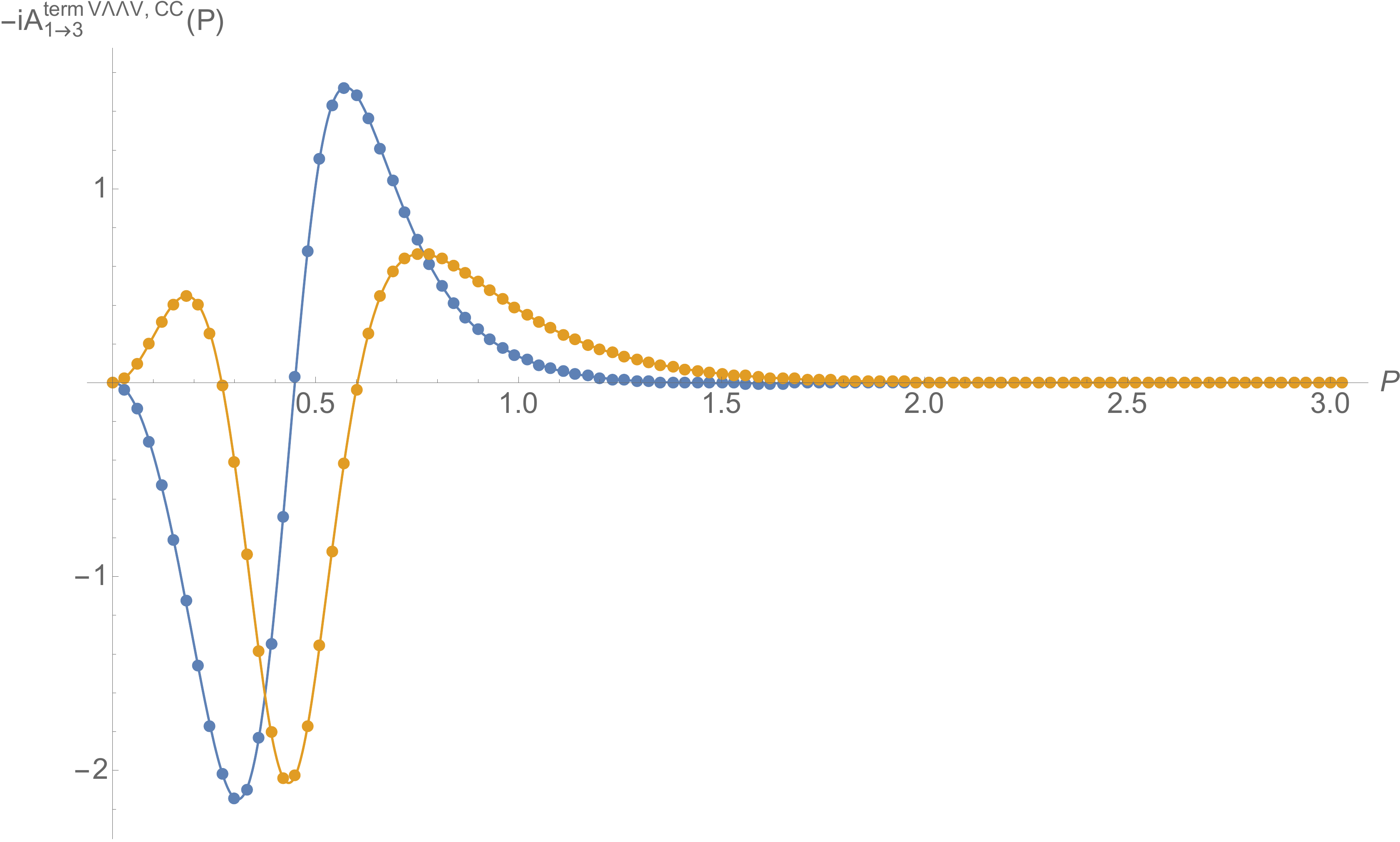} &~~~& \includegraphics[width=0.4\textwidth]{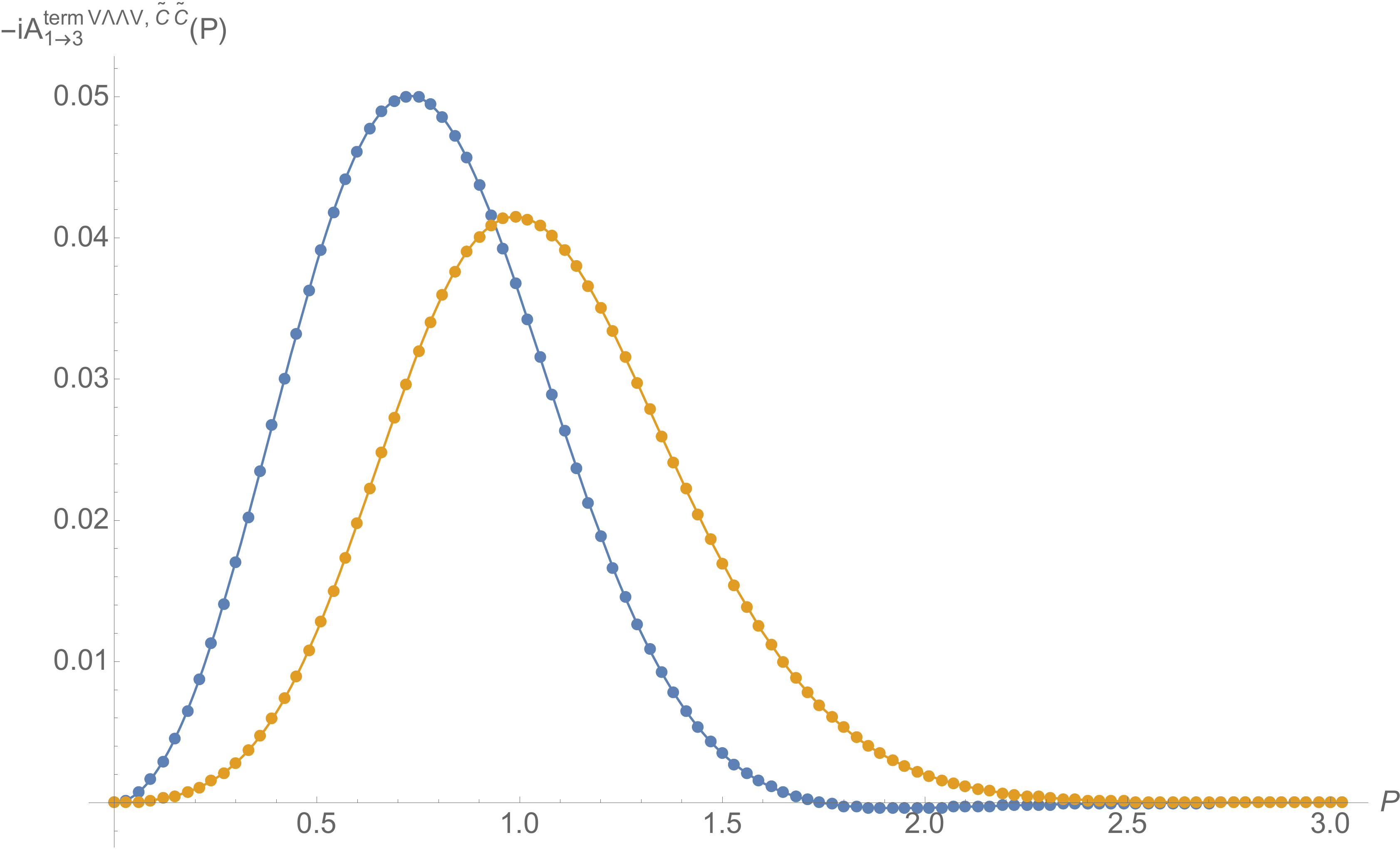}
\end{tabular}
\caption{Sample plots of the $P$-integrand of the various terms in the regularized amplitude (\ref{eq:1to3ampreg}) with the choice (\ref{eq:choice1}) and $\text{Im}(\omega)=0.6$ for the external energies, after having performed the $z$-integral following the strategy outlined above with $\epsilon=10^{-2}$.
}
\label{samplePplots}
\end{figure}

For a direct numerical test of the regularized tree-level $1_R\to 3_R$ amplitude (\ref{eq:1to3ampreg}) against the MQM result, we will make the following choice of incoming and outgoing closed string energies:
\ie
& \omega = \frac{1}{3} + i a,~~~ a\in [0.5, 0.7], 
\\
& \omega_1 = \omega_2 = \omega_3 = \frac{\omega}{3}.
\label{eq:choice1}
\fe
In particular, it is convenient to choose suitably large values of of ${\rm Im}(\omega)$ so as to minimize the number of regulator counterterms in (\ref{eq:1to3ampreg}). In analytically continuing $\omega$ from the positive real axis to the range considered in (\ref{eq:choice1}), no poles in the relevant structure constants cross the integration contour in the Liouville momentum $P$. 
In this case, the only nonzero regulators are $R_t^{\scriptscriptstyle VVVV}, R_t^{\scriptscriptstyle VWWV},$ and $R_t^{\scriptscriptstyle V\Lambda\Lambda V}$, and in these the only terms that contribute are the leading order ones involving the NS sector $C$ structure constants.
%Further numerical details of this computation are presented in Appendix \ref{numerics}. 
In practice, it is convenient to first perform the integration in $z$ with the strategy outlined in the preceding paragraph, at a given value of the internal Liouville momentum $P$, and integrate over $P$ in the end. Sample plots of the $P$-integrand are shown in Figure \ref{samplePplots}.

\begin{figure}[h!]
\centering
\includegraphics[width=0.85\textwidth]{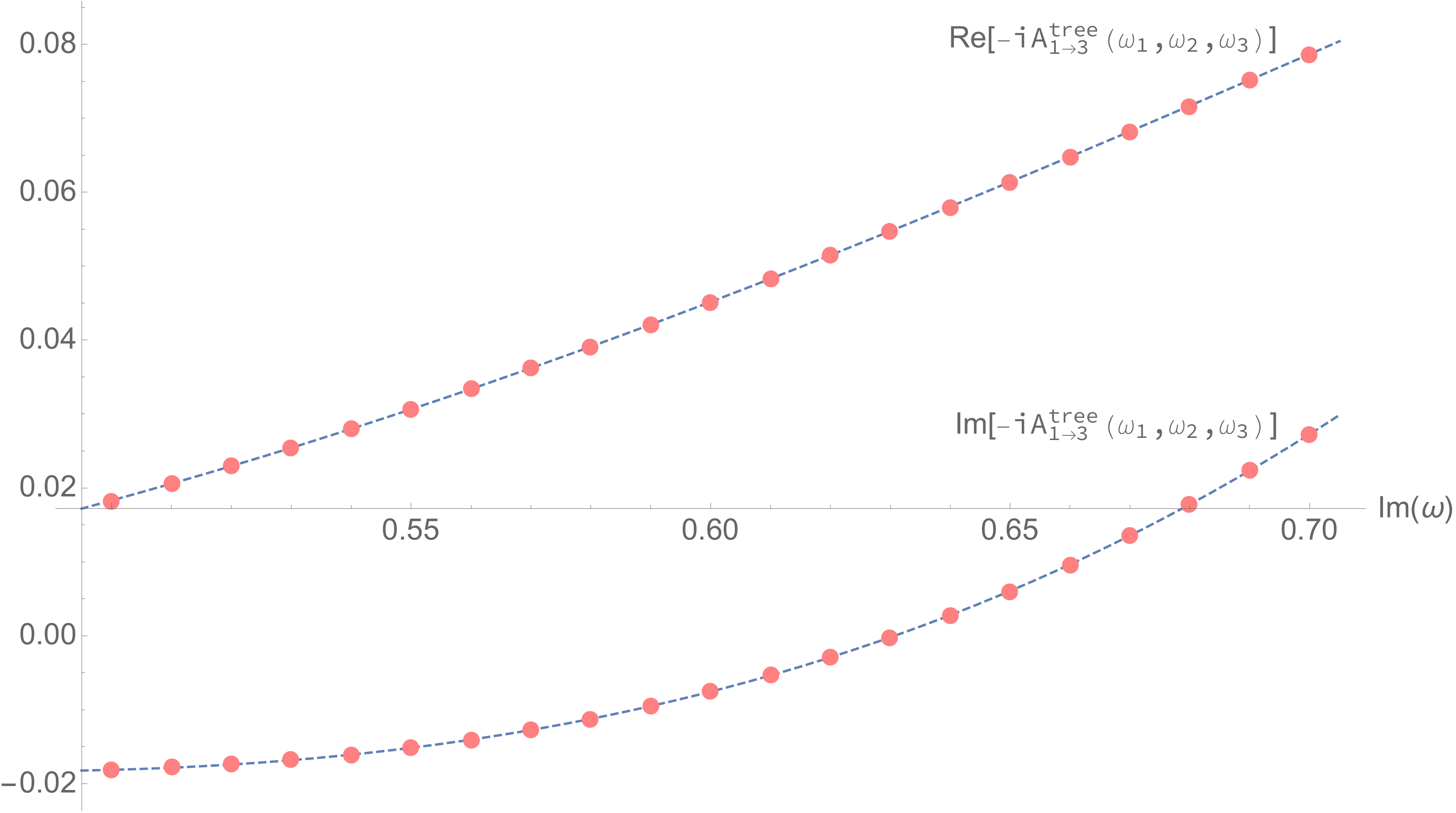}
\caption{Shown in red dots are the numerical results for the real and imaginary parts of the $1_R\to 3_R$ amplitude of closed strings, given by equation (\ref{eq:1to3ampreg}) with the energy assignment (\ref{eq:choice1}) for the incoming and outgoing closed strings. The anticipated matrix model result is shown in blue dashed lines.}
\label{fig:1to3ampWSvsMM}
\end{figure}

The results of the numerical evaluation of the regularized amplitude (\ref{eq:1to3ampreg}) at complex energies (\ref{eq:choice1}), computed following the integration strategy outlined above with $\epsilon=10^{-2}$, are shown in Figure \ref{fig:1to3ampWSvsMM} to be in good agreement with the MQM result (\ref{eq:pertAmpsMM}). %The deviation between the two results is less than $0.3\%$ (not visible in the plot), within the expected numerical error \fixme{fix} due to the numerical interpolation and integration in the Liouville momenta and the cross ratio (and to a lesser extent, the truncation of conformal blocks).
The deviation between the worldsheet computation and the MQM result is less than $0.3\%$ (not visible in the plot). The main source of this deviation is the low truncation order of the superconformal blocks used near $z=1$. Another source of error, to a lesser extent, is the numerical interpolation needed to perform the integration in the cross ratio $z$ and in the internal Liouville momentum $P$.

\section{Discussion}
\label{sec:discuss}

In this paper, we have carefully identified the vertex operators for closed string asymptotic states in the worldsheet description of the two-dimensional type 0B string theory, and performed a direct computation of tree-level scattering amplitudes. At the level of $1\to 2$ amplitudes, we verified the perturbative decoupling between the modes dual to the collective excitations of the ``left" and ``right" fermi sea. Much more nontrivially, we analyzed the $1\to 3$ amplitude and numerically verified its agreement with the conjectured dual matrix quantum mechanics \cite{Douglas:2003up,Takayanagi:2003sm}. 

The key technical ingredient employed and developed in this paper is the recurrence relation of ${\cal N}=1$ super-Virasoro conformal blocks with NS sector external states. Though not needed for the computations performed in this paper, the $1 \to 3$ amplitude of type 0B string theory provides a good target in the near future for formulating and implementing analogous recursion relations for the super-Virasoro conformal blocks involving R sector states. Furthermore, it would be desirable to extend the analysis of \cite{Cho:2017oxl} to derive recursion formulae for super-Virasoro conformal blocks on higher-genus Riemann surfaces, which would allow for a direct verification of the modular invariance of ${\cal N}=1$ Liouville theory and the computation of loop amplitudes in type 0B string theory. The latter may serve as a testing ground for the prescription of superstring perturbation theory based on PCOs and vertical integration that may be necessary in handling spurious singularities \cite{Sen:2014pia, Sen:2015hia}. More broadly, these recursion formulae will be useful for the numerical bootstrap of 2D ${\cal N}=1$ superconformal theories.

Another potential application of the tools developed in this paper is the two-dimensional heterotic $SO(23)$ string theory \cite{Davis:2005qe}, whose worldsheet CFT also involves $\cN=1$ Liouville theory at $b=1$. It is not known whether a MQM dual exists in this case. Useful hints may be provided by the evaluation of perturbative amplitudes similar to the ones performed here.
%As a broader application, one can use the recursion relations written in this paper in order to perform $\cN=1$ super-Virasoro conformal bootstrap in the NS sector. 

It would also be interesting to extend the methods developed in this paper to study scattering amplitudes of open strings in type 0B string theory. 
At leading order, computing these amplitudes requires combining the boundary structure constants of the ${\cal N}=1$ Liouville theory bootstrapped in \cite{Fukuda:2002bv} and the recursion relations for superconformal blocks described in this work. It would be particularly interesting to study the ``long string limit" of open strings ending on FZZT branes, which have been conjectured to be dual to states in non-singlet sectors of the dual matrix model \cite{Maldacena:2005hi}. A non-trivial check of this proposal is to match worldsheet scattering amplitudes with the corresponding amplitudes predicted from the (ungauged) matrix model,  as was done in the bosonic $c=1$ string theory \cite{Balthazar:2018qdv}.

A main motivation for our revisitation of the 2D type 0B string theory is to analyze its non-perturbative sector. Having established the conventions of perturbation theory and verified its validity, we will study the non-perturbative effects mediated by D-instantons in a companion paper \cite{paper:partii}.

%%%%%
\section*{Acknowledgements}

%We would like to thank Nathan Agmon, Alexander Belavin, Panos Betzios, Minjae Cho, Igor Klebanov, Juan Maldacena, and Ashoke Sen for discussions. 
We would like to thank Alexander Belavin for discussions. 
We are especially grateful to Ying-Hsuan Lin for sharing his code implementing a brute force calculation of super-Virasoro conformal blocks in verifying the recurrence relations. XY thanks Cargese Summer Institute, Aspen Center for Physics, Massachusetts Institute of Technology, VR and XY thank Kavli Institute for Theoretical Physics, for their hospitality during the course of this work.
This work is supported in part by a Simons Investigator Award from the Simons Foundation, by the Simons Collaboration Grant on the Non-Perturbative Bootstrap, and by DOE grants DE-SC0007870 and DE-SC0009924.

%%%%%%%%%% APPENDIX %%%%%%%%%%
\appendix

\section{Crossing invariance in $\cN=1$ Liouville theory}
\label{crossing}

Here we present a numerical verification of the crossing invariance of sphere four-point functions of NS sector states in $\cN=1$ Liouville theory, which serves as a highly nontrivial consistency check of both the structure constants (\ref{eq:superDOZZNS}) and the $c$-recusion formulae for the superconformal blocks given in section \ref{crecursions}.

The four-point functions that enter the moduli integrand of the $1\to 3$ amplitude of type 0B strings, along with their superconformal block decompositions, are
\ie
&G(4321|z) := \langle V_{P_4}(\infty)V_{P_3}(1)V_{P_2}(z,\zbar)V_{P_1}(0) \rangle \vphantom{\frac{1}{1}}\\
& = \int_0^\infty \frac{dP}{\pi} \left[ C(P_1,P_2,P)C(P_3,P_4,P) \cF^e(h_4,h_3,h_2,h_1;h_P|z)\cF^e(h_4,h_3,h_2,h_1;h_P|\zbar) \vphantom{\frac{1}{1}}\right. \\
& ~~~~~~~~~~~~ + \left. \widetilde{C}(P_1,P_2,P)\widetilde{C}(P_3,P_4,P) \cF^{o}(h_4,h_3,h_2,h_1;h_P|z)\cF^{o}(h_4,h_3,h_2,h_1;h_P|\zbar) \vphantom{\frac{1}{1}}\right],
\label{eq:LiouvVVVV}
\fe
\ie
& H(4321|z) := \langle V_{P_4}(\infty)W_{P_3}(1)W_{P_2}(z,\zbar)V_{P_1}(0) \rangle \vphantom{\frac{1}{1}}\\
& = -\int_0^\infty \frac{dP}{\pi} \left[ \widetilde{C}(P_1,P_2,P)\widetilde{C}(P_3,P_4,P) \cF^e(h_4,*h_3,*h_2,h_1;h_P|z)\cF^e(h_4,*h_3,*h_2,h_1;h_P|\zbar) \vphantom{\frac{1}{1}}\right. \\
& ~~~~~~~~~~~~~~~ + \left. {C}(P_1,P_2,P){C}(P_3,P_4,P) \cF^{o}(h_4,*h_3,*h_2,h_1;h_P|z)\cF^{o}(h_4,*h_3,*h_2,h_1;h_P|\zbar) \vphantom{\frac{1}{1}}\right],
\label{eq:LiouvVWWV}
\fe
where $h_i = \frac{1}{2}(1+P_i^2)$ and $h_P = \frac{1}{2}(1+P^2)$ for ${\cal N}=1$ Liouville theory at $b=1$, as well as
\ie
J(&4321|z) := \langle V_{P_4}(\infty)\widetilde{\psi}^0\Lambda_{P_3}(1)\widetilde{\psi}^0\Lambda_{P_2}(z,\zbar)V_{P_1}(0) \rangle + \langle V_{P_4}(\infty){\psi}^0\widetilde{\Lambda}_{P_3}(1){\psi}^0\widetilde{\Lambda}_{P_2}(z,\zbar)V_{P_1}(0) \rangle \vphantom{\frac{1}{1}}\\
= & -\int_0^\infty \frac{dP}{\pi} \left[ C(P_1,P_2,P)C(P_3,P_4,P) \left(\frac{1}{1-\zbar}\cF^{o}(h_4,*h_3,*h_2,h_1;h_P|z)\cF^e(h_4,h_3,h_2,h_1;h_P|\zbar) \right.\right. \\
& ~~~~~~~~~~~~~~~~~~~~~~~~~~~~~~~~~~~~~~~~~~~~~~~~ \left. + \frac{1}{1-z}\cF^e(h_4,h_3,h_2,h_1;h_P|z) \cF^{o}(h_4,*h_3,*h_2,h_1;h_P|\zbar) \right) \\
& ~~~~~~ \left. +\,\, \widetilde{C}(P_1,P_2,P)\widetilde{C}(P_3,P_4,P) \left(\frac{1}{1-\zbar}\cF^e(h_4,*h_3,*h_2,h_1;h_P|z)\cF^{o}(h_4,h_3,h_2,h_1;h_P|\zbar) \right.\right. \\
& ~~~~~~~~~~~~~~~~~~~~~~~~~~~~~~~~~~~~~~~~~~ \left.\left. + \frac{1}{1-z}\cF^{o}(h_4,h_3,h_2,h_1;h_P|z) \cF^e(h_4,*h_3,*h_2,h_1;h_P|\zbar) \right)\right].
\label{eq:LiouvVLLV}
\fe
In (\ref{eq:LiouvVLLV}), for convenience we included the time-like free fermion (which appears in the pictured-raised tachyon vertex operator (\ref{eq:tachraised})) in the correlator so that the operators $\widetilde{\psi}^0\Lambda_P$ and $\psi^0 \widetilde{\Lambda}_P$ are scalars of weight $h=\widetilde{h}=h_P + \frac{1}{2}$. %Furthermore, we added both such type of contributions together in (\ref{eq:tachraised}) 
Note that the RHS of (\ref{eq:LiouvVWWV}) and (\ref{eq:LiouvVLLV}) are manifestly real and negative due to the anti-Hermiticity of the operators $W_P$, $\widetilde{\psi}^0\Lambda_{P}$, and ${\psi}^0\widetilde{\Lambda}_{P}$. %\footnote{Note that the integrands in square brackets of (\ref{eq:LiouvVVVV}), (\ref{eq:LiouvVWWV}) and (\ref{eq:LiouvVLLV}) are real-valued and positive for $P_i,P\in \bR_{+}$.}

\begin{figure}[h!]
\centering
\begin{tabular}{c}
\includegraphics[width=0.5\textwidth]{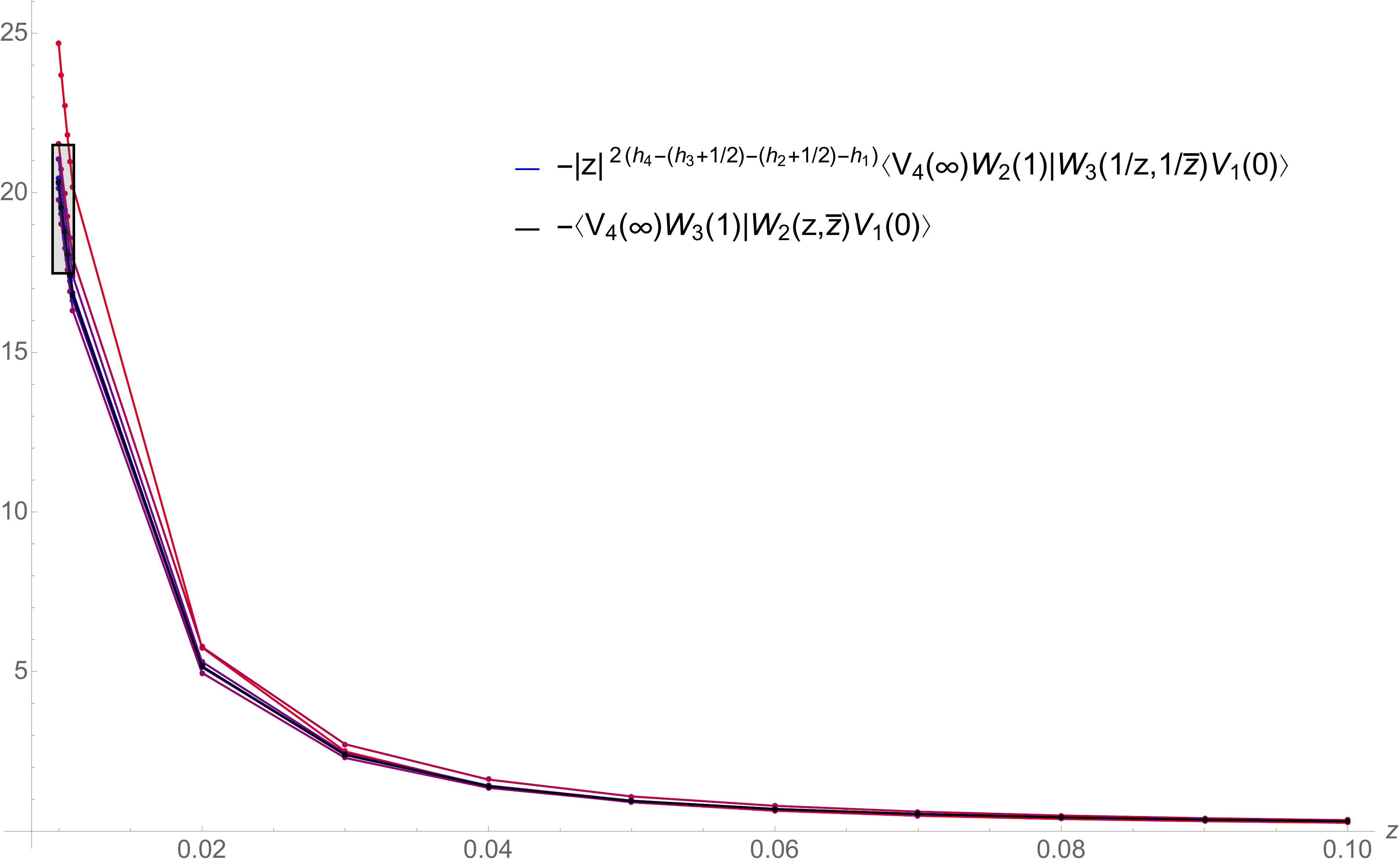}~~
 \includegraphics[width=0.5\textwidth]{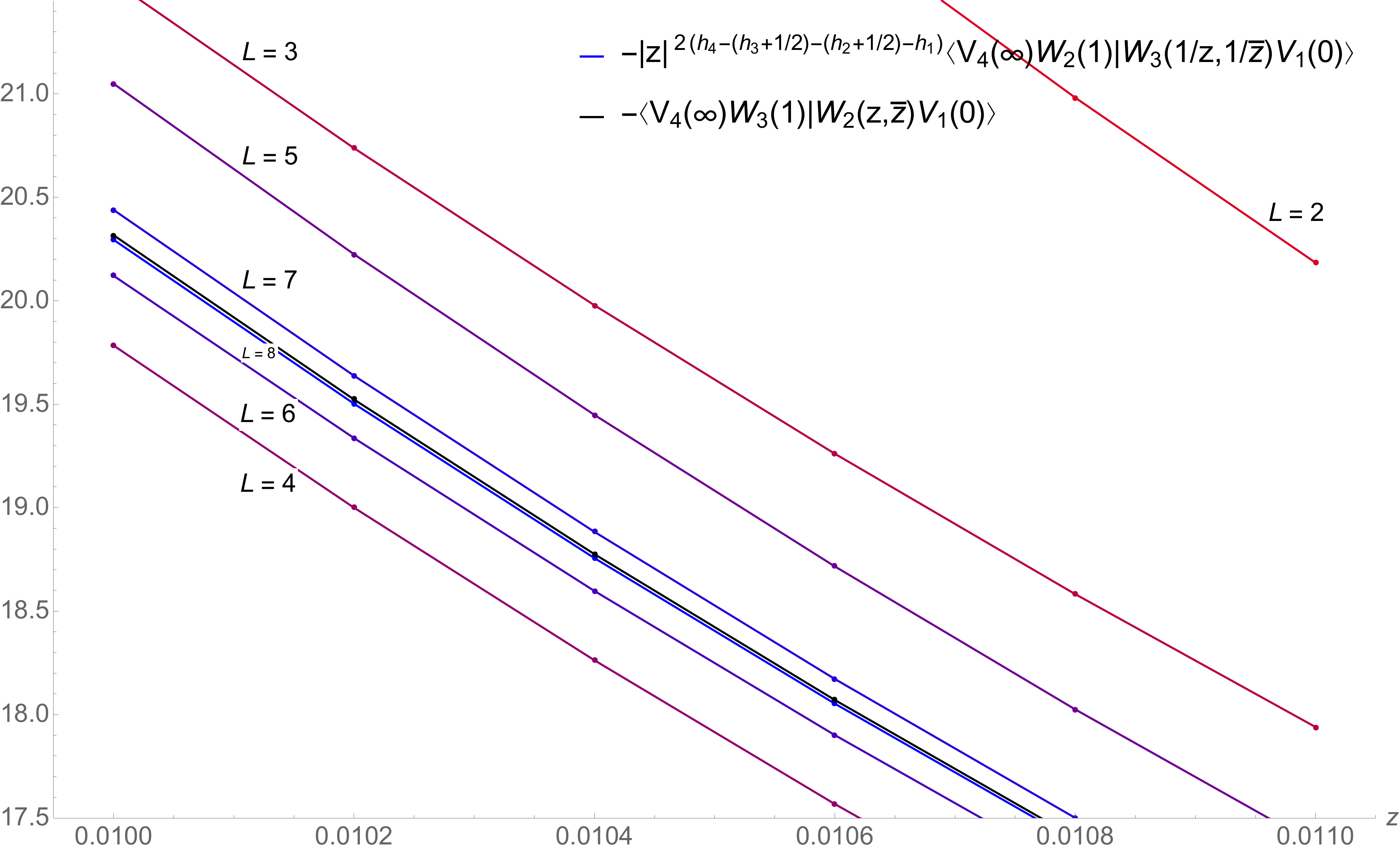}
\end{tabular}
\caption{Numerical test of the crossing invarince of $-\langle V_{P_4}(\infty)W_{P_3}(1)W_{P_2}(z,\zbar)V_{P_1}(0) \rangle$ over a range of small positive real $z$, at (arbitrarily chosen) external Liouville momenta $(P_1, P_2, P_3, P_4)=({1\over 2},{1\over 3},{1\over 4},{3\over 5})$. The crossed channel (RHS of (\ref{eq:Scrossing})) computed with the superconformal blocks truncated to order $q^{L}$ is shown in the color scheme from red to blue for $L$ increasing from 2 to 8. The direct channel (LHS of (\ref{eq:Scrossing})) result computed using the superconformal block expansion truncated to order $q^8$ is shown in black. The right figure is a zoomed-in view of the boxed region in the left figure. The data points are joined with straight lines for visualization.}
\label{fig:VWWVcrossing}
\end{figure}

The crossing relations that amount to exchanging the roles of the operators $2\leftrightarrow 3$ in the OPE in the superconformal block decomposition of (\ref{eq:4pt}) read
\ie
G(4321|z) &= |z|^{2(h_4-h_3-h_2-h_1)}G(4231|z^{-1}), \\
H(4321|z) &= |z|^{2(h_4-(h_3+1/2)-(h_2+1/2)-h_1)}H(4231|z^{-1}), \\
J(4321|z) &= |z|^{2(h_4-(h_3+1/2)-(h_2+1/2)-h_1)}J(4231|z^{-1}).
\label{eq:Scrossing}
\fe
%Although not necessary for the computations in this paper of string amplitudes in type 0B string theory, we have also verified crossing symmetry involving superconformal blocks of type $VVWV$ and $VWVV$, which reads
%\ie
%\langle V_{P_4}(\infty)W_{P_3}(1)V_{P_2}(z,\zbar)V_{P_1}(0) \rangle &= |z|^{2(h_4-(h_3+1/2)-h_2-h_1)} \langle V_{P_4}(\infty)V_{P_2}(1)W_{P_3}(z^{-1},\zbar^{-1})V_{P_1}(0) \rangle,
%\label{eq:VWVVcrossing}
%\fe
%or expanding in superconformal blocks,
%\ie
%&\int_0^\infty \frac{dP}{\pi} \left[ {C}(P_1,P_2,P)\widetilde{C}(P_3,P_4,P) \cF^e(h_4,*h_3,h_2,h_1;h_P|z)\cF^e(h_4,*h_3,h_2,h_1;h_P|\zbar) \vphantom{\frac{1}{1}}\right. \\
%& ~~~~~~~~~ + \left. \widetilde{C}(P_1,P_2,P){C}(P_3,P_4,P) \cF^{o}(h_4,*h_3,h_2,h_1;h_P|z)\cF^{o}(h_4,*h_3,h_2,h_1;h_P|\zbar) \vphantom{\frac{1}{1}}\right]\\
%& = |z|^{2(h_4-(h_3+1/2)-h_2-h_1)} \vphantom{\frac{1}{1}}\\
%& ~~~ \int_0^\infty \frac{dP}{\pi} \left[ \widetilde{C}(P_1,P_3,P)C(P_2,P_4,P) \cF^e(h_4,h_2,*h_3,h_1;h_P|1/z)\cF^e(h_4,h_2,*h_3,h_1;h_P|1/\zbar) \vphantom{\frac{1}{1}}\right. \\
%& ~~~~~~~~~~~~ + \left. {C}(P_1,P_3,P)\widetilde{C}(P_2,P_4,P) \cF^{o}(h_4,h_2,*h_3,h_1;h_P|1/z)\cF^{o}(h_4,h_2,*h_3,h_1;h_P|1/\zbar) \vphantom{\frac{1}{1}}\right]. \\
%\fe
A sample numerical verification of the crossing invariance, of the correlator of type $\langle VWWV\rangle$ (second line of (\ref{eq:Scrossing})), is shown in Figure \ref{fig:VWWVcrossing}. For a range of small positive real $z$, the ``crossed" $13\to 24$ channel result computed with increasing truncation order (red to blue) in its expansion in the elliptic nome $q$ is seen to converge to the ``direct" $12\to 34$ channel result (black).

There are also crossing relations exchanging operators $1\leftrightarrow 3$ that read
%\begin{flalign} 
%\langle V_{P_4}(\infty)V_{P_3}(1)V_{P_2}(z,\zbar)V_{P_1}(0) \rangle = \langle V_{P_4}(\infty)V_{P_1}(1)V_{P_2}(1-z,1-\zbar)V_{P_3}(0) \rangle, && 
%\label{eq:TcrossingVVVV}
%\end{flalign} 
\ie
\langle & V_{P_4}(\infty)W_{P_3}(1)W_{P_2}(z,\zbar)V_{P_1}(0) \rangle \vphantom{\frac{1}{1}} \\
& =  \langle V_{P_4}(\infty)V_{P_1}(1)W_{P_2}(1-z,1-\zbar)W_{P_3}(0) \rangle \vphantom{\frac{1}{1}} \\
& =  -\int_0^\infty \frac{dP}{\pi} \left[ {C}(P_1,P_4,P){C}(P_2,P_3,P) \cF^e(h_4,h_1,*h_2,*h_3;h_P|1-z)\cF^e(h_4,h_1,*h_2,*h_3;h_P|1-\zbar) \vphantom{\frac{1}{1}}\right. \\
& ~~~~~~~~~~~~~~~ + \left. \widetilde{C}(P_1,P_4,P)\widetilde{C}(P_2,P_3,P) \cF^{o}(h_4,h_1,*h_2,*h_3;h_P|1-z)\cF^{o}(h_4,h_1,*h_2,*h_3;h_P|1-\zbar) \vphantom{\frac{1}{1}}\right],
\label{eq:TcrossingVWWV}
\fe
and
\ie
\langle & V_{P_4}(\infty)\widetilde{\psi}^0\Lambda_{P_3}(1)\widetilde{\psi}^0\Lambda_{P_2}(z,\zbar)V_{P_1}(0) \rangle + \langle V_{P_4}(\infty){\psi}^0\widetilde{\Lambda}_{P_3}(1){\psi}^0\widetilde{\Lambda}_{P_2}(z,\zbar)V_{P_1}(0) \rangle \vphantom{\frac{1}{1}}\\
& = \langle V_{P_4}(\infty)V_{P_1}(1)\widetilde{\psi}^0\Lambda_{P_2}(1-z,1-\zbar) \widetilde{\psi}^0\Lambda_{P_3}(0) \rangle + \langle V_{P_4}(\infty)V_{P_1}(0){\psi}^0\widetilde{\Lambda}_{P_2}(1-z,1-\zbar){\psi}^0\widetilde{\Lambda}_{P_3}(0) \rangle \vphantom{\frac{1}{1}}\\
& = -\int_0^\infty \frac{dP}{\pi} \left[ C(P_1,P_4,P)C(P_2,P_3,P) \left(\frac{1}{1-\zbar}\cF^e(h_4,h_1,*h_2,*h_3;h_P|1-z)\cF^e(h_4,h_1,h_2,h_3;h_P|1-\zbar) \right.\right. \\
& ~~~~~~~~~~~~~~~~~~~~~~~~~~~~~~~~~~~~~~~~~~~~~~~~~~~ \left. + \frac{1}{1-z}\cF^e(h_4,h_1,h_2,h_3;h_P|1-z)\cF^e(h_4,h_1,*h_2,*h_3;h_P|1-\zbar) \right) \\
& ~~~~~~ \left. +\,\, \widetilde{C}(P_1,P_4,P)\widetilde{C}(P_2,P_3,P) \left(\frac{1}{1-\zbar}\cF^{o}(h_4,h_1,*h_2,*h_3;h_P|1-z)\cF^{o}(h_4,h_1,h_2,h_3;h_P|1-\zbar) \right.\right. \\
& ~~~~~~~~~~~~~~~~~~~~~~~~~~~~~~~~~~~~~~~~~~ \left.\left. + \frac{1}{1-z}\cF^{o}(h_4,h_1,h_2,h_3;h_P|1-z)\cF^{o}(h_4,h_1,*h_2,*h_3;h_P|1-\zbar) \right) \right].
\label{eq:TcrossingVLLV}
\fe
In section \ref{crecursions} we did not present recursion relations for super-Virasoro conformal blocks of type $\cF^{e/o}(h_4,h_3,*h_2,*h_1;h_P;z)$ appearing on the RHS of (\ref{eq:TcrossingVWWV}) and (\ref{eq:TcrossingVLLV}). 
These blocks are related to the ones considered in section \ref{crecursions} by superconformal Ward identities. For the purpose of evaluating 4-point type 0B string amplitude in this paper, we will only make use of the decomposition (\ref{eq:TcrossingVWWV}) and (\ref{eq:TcrossingVLLV}) in a small neighborhood around $z=1$, where a truncation to leading or next-to-leading order in $(1-z)$ suffices. Namely,
\ie
\cF^e(h_4,h_1,*h_2,*h_3;h|1-z) & = (1-z)^{h-(h_2+{1\over 2})-(h_3+{1\over 2})} \sum_{n=0}^{\infty} c_n\,(1-z)^n, \\
\cF^{o}(h_4,h_1,*h_2,*h_3;h|1-z) & = (1-z)^{h-(h_2+{1\over 2})-(h_3+{1\over 2})} \sum_{n=0}^{\infty} c_{n+{1\over 2}}\,(1-z)^{n+{1\over 2}},
\label{eq:coeffcpdef}
\fe
where we will only keep the terms with coefficients
\ie
c_0 &= h_2 + h_3 - h, ~~~ c_1 = -\frac{1}{2h}(h - h_3 + h_2)(h + h_1 - h_4)(h - h_2 - h_3), \\
c_{1\over 2} &= \frac{1}{2h}(h+h_2+h_3-{1\over 2}).
\label{eq:coeffcp}
\fe
%For completeness, we list the first few coefficients in a small $z$ expansion of the following blocks,
%\ie
%\cF^e(h_4,h_3,h_2,h_1;h|z) & = z^{h-h_1-h_2} \sum_{n=0}^{\infty} c_n\,z^n \\
%\cF^{o}(h_4,h_3,h_2,h_1;h|z) & = z^{h-h_1-h_2} z^{1/2}\sum_{n=0}^{\infty} c_{n+1/2}\,z^n
%\label{eq:coeffcdef}
%\fe
%\ie
%\cF^e(h_4,*h_3,*h_2,h_1;h|z) & = z^{h-h_1-(h_2+1/2)} \sum_{n=0}^{\infty} \widetilde{c}_n\,z^n \\
%\cF^{o}(h_4,*h_3,*h_2,h_1;h|z) & = z^{h-h_1-(h_2+1/2)} z^{1/2}\sum_{n=0}^{\infty} \widetilde{c}_{n+1/2}\,z^n
%\label{eq:coeffctdef}
%\fe
%\ie
%&c_0 = 1, ~~~ c_1 = \frac{1}{2h}(h-h_1+h_2)(h+h_3-h_4), ~~~ c_{1/2} = \frac{1}{2h}, \\
%&c_{3/2} =\frac{(1/2+h - h_1 + h_2)(1/2 + h + h_3 - h_4)}{2h(1+2h)}-\frac{6(h_1-h_2)(h_3-h_4)}{(1 + 2 h) (c + 2 c h + 3 h (-3 + 2 h))},
%\label{eq:coeffc}
%\fe
%\ie
%& \widetilde{c}_0 = 1, ~~~ \widetilde{c}_1 = \frac{(1/2 + h - h_1 + h_2) (1/2 + h + h_3 - h_4)}{2h}, ~~~ \widetilde{c}_{1/2} = \frac{(h - h_1 + h_2) (h + h_3 - h_4)}{2h},\\
%&\widetilde{c}_{3/2} = \frac{(h - h_1 + h_2) (1 + h - h_1 + h_2) (h + h_3 - h_4) (1 + h + h_3 - h_4)}{2 h (1 + 2 h)} \\
%& + \frac{3 (h_1 - 2 h_1^2 + h_2 + 4 h_1 h_2 - 2 h_2^2 + h (2 h_1 + 2 h_2 - 1 )) (h_3 - 
%   2 h_3^2 + h_4 + 4 h_3 h_4 - 2 h_4^2 + h ( 2 h_3 + 2 h_4 -1  ))}{2 (1 + 2 h) (c + 2 c h + 3 h (-3 + 2 h))},
%\label{eq:coeffct}
%\fe
With such low order approximations to the blocks, we find (\ref{eq:TcrossingVWWV}) and (\ref{eq:TcrossingVLLV}) to be in good agreement with the direct channel result at $z$ close to 1.

We have also performed similar tests of crossing invariance for four-point functions in ${\cal N}=1$ Liouville theory at more general values of $b$, although only the $b=1$ case is relevant for the worldsheet theory of the 2D type 0B string.

\section{Regularizing the moduli integration}
\label{regularization}

The expression for the $1\to 3$ amplitude (\ref{eq:1to3ampnaive}) involves a moduli integral that a priori diverges when pairs of vertex operators collide. Closely following the treatment of \cite{Balthazar:2017mxh}, we regularize the moduli integral by explicitly subtracting counter terms that cancel the power divergences. %In type 0B string theory, whose worldsheet CFT is supersymmetric, there are more types of conformal blocks (c.f. (\ref{eq:sCBs})) and hence a larger number of regulator counterterms that we need to include. 

Near $z=0$, the $z$-integrand of (\ref{eq:1to3ampnaive}) can be expanded as
\ie{}
&\omega_2^2\omega_3^2 \int_0^{\infty}\frac{dP}{\pi} \left[ C(\omega_1,\omega_2,P)C(\omega_3,\omega,P) |z|^{-1+P^2-(\omega_1+\omega_2)^2} \sum_{n,m\in\bZ^{\geq 0}} a^{\scriptscriptstyle{VVVV}}_{n,m}z^n \overline{z}^m \right.
\\
&~~~~~~~~\left. + \widetilde{C}(\omega_1,\omega_2,P)\widetilde{C}(\omega_3,\omega,P) |z|^{P^2-(\omega_1+\omega_2)^2} \sum_{n,m\in\bZ^{\geq 0}} a'^{\scriptscriptstyle{VVVV}}_{n,m}z^n \overline{z}^m \right]
\\
&+\int_0^{\infty}\frac{dP}{\pi} \left[ \widetilde{C}(\omega_1,\omega_2,P)\widetilde{C}(\omega_3,\omega,P) |z|^{-2+P^2-(\omega_1+\omega_2)^2} \sum_{n,m\in\bZ^{\geq 0}} a^{\scriptscriptstyle{VWWV}}_{n,m}z^n \overline{z}^m \right.
\\
&~~~~~~~~\left. + C(\omega_1,\omega_2,P)C(\omega_3,\omega,P) |z|^{-1+P^2-(\omega_1+\omega_2)^2} \sum_{n,m\in\bZ^{\geq 0}} a'^{\scriptscriptstyle{VWWV}}_{n,m}z^n \overline{z}^m \right]
\\
&+\omega_2\omega_3 \int_0^{\infty}\frac{dP}{\pi} \left[ C(\omega_1,\omega_2,P)C(\omega_3,\omega,P) |z|^{-1+P^2-(\omega_1+\omega_2)^2} \sum_{n,m\in\bZ^{\geq 0}} a^{\scriptscriptstyle{V\Lambda\Lambda V}}_{n,m}z^n \overline{z}^m \right.
\\
&~~~~~~~~\left. + \widetilde{C}(\omega_1,\omega_2,P)\widetilde{C}(\omega_3,\omega,P) |z|^{-2+P^2-(\omega_1+\omega_2)^2} \sum_{n,m\in\bZ^{\geq 0}} a'^{\scriptscriptstyle{V\Lambda\Lambda V}}_{n,m}z^n \overline{z}^m \right],
\label{eq:expansionz0}
\fe
where the coefficients $a^{('){\scriptscriptstyle V\_ \_ V}}_{m,n}$ can be calculated using the superconformal blocks appearing in the decomposition of $\cN=1$ Liouville correlators. %\footnote{See Appendix \ref{crossing} for explicit expressions of low order coefficients.} 
The $z$-integral diverges when $P^2 - \text{Re}[(\omega_1+\omega_2)^2]< 0$, $-1$, or $-2$ for the various terms appearing in (\ref{eq:expansionz0}). The moduli integrand near $z=\infty$ takes precisely the same form as (\ref{eq:expansionz0}) upon exchanging $\omega_2\leftrightarrow\omega_3$ and $z\leftrightarrow z^{-1}$ (c.f. (\ref{eq:Scrossing})).

Near $z=1$, on the other hand, the integrand takes the form
\ie{}
&\omega_2^2\omega_3^2\int_1 d^2z\int_0^{\infty}\frac{dP}{\pi} \left[ C(\omega_1,\omega,P)C(\omega_2,\omega_3,P) |1-z|^{-3+P^2-(\omega_2+\omega_3)^2} \sum_{n,m\in\bZ^{\geq 0}} b^{\scriptscriptstyle{VVVV}}_{n,m}(1-z)^n (1-\overline{z})^m \right.
\\
&~~~~~~~~\left. + \widetilde{C}(\omega_1,\omega,P)\widetilde{C}(\omega_2,\omega_3,P) |1-z|^{-2+P^2-(\omega_2+\omega_3)^2} \sum_{n,m\in\bZ^{\geq 0}} b'^{\scriptscriptstyle{VVVV}}_{n,m}(1-z)^n (1-\overline{z})^m \right]
\\
&+\int_1 d^2z\int_0^{\infty}\frac{dP}{\pi} \left[ C(\omega_1,\omega,P)C(\omega_2,\omega_3,P) |1-z|^{-3+P^2-(\omega_2+\omega_3)^2} \sum_{n,m\in\bZ^{\geq 0}} b^{\scriptscriptstyle{VWWV}}_{n,m}(1-z)^n (1-\overline{z})^m \right.
\\
&~~~~~~~~\left. + \widetilde{C}(\omega_1,\omega,P)\widetilde{C}(\omega_2,\omega_3,P) |1-z|^{-2+P^2-(\omega_2+\omega_3)^2} \sum_{n,m\in\bZ^{\geq 0}} b'^{\scriptscriptstyle{VWWV}}_{n,m}(1-z)^n (1-\overline{z})^m \right]
\\
&+\omega_2\omega_3\int_1 d^2z\int_0^{\infty}\frac{dP}{\pi} \left[ C(\omega_1,\omega,P)C(\omega_2,\omega_3,P) |1-z|^{-3+P^2-(\omega_2+\omega_3)^2} \sum_{n,m\in\bZ^{\geq 0}} b^{\scriptscriptstyle{V\Lambda\Lambda V}}_{n,m}(1-z)^n (1-\overline{z})^m \right.
\\
&~~~~~~~~\left. + \widetilde{C}(\omega_1,\omega,P)\widetilde{C}(\omega_2,\omega_3,P) |1-z|^{-2+P^2-(\omega_2+\omega_3)^2} \sum_{n,m\in\bZ^{\geq 0}} b'^{\scriptscriptstyle{V\Lambda\Lambda V}}_{n,m}(1-z)^n (1-\overline{z})^m \right].
\label{eq:expansionz1}
\fe
The various term give rise to a divergence of the $z$-integral when $P^2-\text{Re}[(\omega_2+\omega_3)^2] < 0$ or 1. Note that the divergence near $z=1$ is of a stronger type than those near $z=0,\infty$. This is due to the spurious singularity from the collision of PCOs, as we have chosen to place the PCOs to coincide with the vertex operator $\cT_{\omega_3}$ at $1$ and $\cT_{\omega_2}$ at $z$. 

The counter terms that serve to remove the power divergences in the $z$-integral are
\ie
R&_s^{\scriptscriptstyle VVVV} = \sum_{0\leq n \leq \frac{1}{2}(-1+\text{Re}[(\omega_1+\omega_2)^2])} a^{\scriptscriptstyle VVVV}_{n,n} \int_0^{\sqrt{-1+\text{Re}[(\omega_1+\omega_2)^2]-2n}}\frac{dP}{\pi} C(\omega_1,\omega_2,P)C(\omega_3,\omega,P) |z|^{-1+P^2-(\omega_1+\omega_2)^2+2n} \\
& + \sum_{0\leq n \leq \frac{1}{2}(-2+\text{Re}[(\omega_1+\omega_2)^2])} a'^{\scriptscriptstyle VVVV}_{n,n} \int_0^{\sqrt{-2+\text{Re}[(\omega_1+\omega_2)^2]-2n}}\frac{dP}{\pi} \widetilde{C}(\omega_1,\omega_2,P)\widetilde{C}(\omega_3,\omega,P) |z|^{P^2-(\omega_1+\omega_2)^2+2n} \\
R&_u^{\scriptscriptstyle VVVV} = |z|^{-4} \left(R_s^{\scriptscriptstyle VVVV}|_{z\to 1/z, \omega_2\leftrightarrow\omega_3}\right) \\
R&_t^{\scriptscriptstyle VVVV} = \sum_{0\leq n \leq \frac{1}{2}(1+\text{Re}[(\omega_2+\omega_3)^2])} b^{\scriptscriptstyle VVVV}_{n,n} \int_0^{\sqrt{1+\text{Re}[(\omega_2+\omega_3)^2]-2n}}\frac{dP}{\pi} C(\omega_1,\omega,P)C(\omega_2,\omega_3,P) |z|^{-3+P^2-(\omega_2+\omega_3)^2+2n}\\
& + \sum_{0\leq n \leq \frac{1}{2}\text{Re}[(\omega_2+\omega_3)^2]} b'^{\scriptscriptstyle VVVV}_{n,n} \int_0^{\sqrt{\text{Re}[(\omega_2+\omega_3)^2]-2n}}\frac{dP}{\pi} \widetilde{C}(\omega_1,\omega,P)\widetilde{C}(\omega_2,\omega_3,P) |z|^{-2+P^2-(\omega_2+\omega_3)^2+2n}
\fe
\ie
R&_s^{\scriptscriptstyle VWWV} = \sum_{0\leq n \leq \frac{1}{2}\text{Re}[(\omega_1+\omega_2)^2]} a^{\scriptscriptstyle VWWV}_{n,n} \int_0^{\sqrt{\text{Re}[(\omega_1+\omega_2)^2]-2n}}\frac{dP}{\pi} \widetilde{C}(\omega_1,\omega_2,P)\widetilde{C}(\omega_3,\omega,P) |z|^{-2+P^2-(\omega_1+\omega_2)^2+2n} \\
& + \sum_{0\leq n \leq \frac{1}{2}(-1+\text{Re}[(\omega_1+\omega_2)^2])} a'^{\scriptscriptstyle VWWV}_{n,n} \int_0^{\sqrt{-1+\text{Re}[(\omega_1+\omega_2)^2]-2n}}\frac{dP}{\pi} C(\omega_1,\omega_2,P)C(\omega_3,\omega,P) |z|^{-1+P^2-(\omega_1+\omega_2)^2+2n} \\
R&_u^{\scriptscriptstyle VWWV} = |z|^{-4} \left(R_s^{\scriptscriptstyle VWWV}|_{z\to 1/z, \omega_2\leftrightarrow\omega_3}\right) \\
R&_t^{\scriptscriptstyle VWWV} = \sum_{0\leq n \leq \frac{1}{2}(1+\text{Re}[(\omega_2+\omega_3)^2])} b^{\scriptscriptstyle VWWV}_{n,n} \int_0^{\sqrt{1+\text{Re}[(\omega_2+\omega_3)^2]-2n}}\frac{dP}{\pi} C(\omega_1,\omega,P)C(\omega_2,\omega_3,P) |z|^{-3+P^2-(\omega_2+\omega_3)^2+2n}\\
& + \sum_{0\leq n \leq \frac{1}{2}\text{Re}[(\omega_2+\omega_3)^2]} b'^{\scriptscriptstyle VWWV}_{n,n} \int_0^{\sqrt{\text{Re}[(\omega_2+\omega_3)^2]-2n}}\frac{dP}{\pi} \widetilde{C}(\omega_1,\omega,P)\widetilde{C}(\omega_2,\omega_3,P) |z|^{-2+P^2-(\omega_2+\omega_3)^2+2n}
\fe
\ie
R&_s^{\scriptscriptstyle V\Lambda\Lambda V} = \sum_{0\leq n \leq \frac{1}{2}(-1+\text{Re}[(\omega_1+\omega_2)^2])} a^{\scriptscriptstyle V\Lambda\Lambda V}_{n,n} \int_0^{\sqrt{-1+\text{Re}[(\omega_1+\omega_2)^2]-2n}}\frac{dP}{\pi} C(\omega_1,\omega_2,P)C(\omega_3,\omega,P) |z|^{-1+P^2-(\omega_1+\omega_2)^2+2n} \\
& + \sum_{0\leq n \leq \frac{1}{2}\text{Re}[(\omega_1+\omega_2)^2]} a^{\scriptscriptstyle V\Lambda\Lambda V}_{n,n} \int_0^{\sqrt{\text{Re}[(\omega_1+\omega_2)^2]-2n}}\frac{dP}{\pi} \widetilde{C}(\omega_1,\omega_2,P)\widetilde{C}(\omega_3,\omega,P) |z|^{-2+P^2-(\omega_1+\omega_2)^2+2n} \\ 
R&_u^{\scriptscriptstyle V\Lambda\Lambda V} = |z|^{-4} \left(R_s^{\scriptscriptstyle V\Lambda\Lambda V}|_{z\to 1/z, \omega_2\leftrightarrow\omega_3}\right) \\
R&_t^{\scriptscriptstyle V\Lambda\Lambda V} =  \sum_{0\leq n \leq \frac{1}{2}(1+\text{Re}[(\omega_2+\omega_3)^2])} b^{\scriptscriptstyle V\Lambda\Lambda V}_{n,n} \int_0^{\sqrt{1+\text{Re}[(\omega_2+\omega_3)^2]-2n}}\frac{dP}{\pi} C(\omega_1,\omega,P)C(\omega_2,\omega_3,P) |z|^{-3+P^2-(\omega_2+\omega_3)^2+2n} \\
& + \sum_{0\leq n \leq \frac{1}{2}\text{Re}[(\omega_2+\omega_3)^2]} b'^{\scriptscriptstyle V\Lambda\Lambda V}_{n,n} \int_0^{\sqrt{\text{Re}[(\omega_2+\omega_3)^2]-2n}}\frac{dP}{\pi} \widetilde{C}(\omega_1,\omega,P)\widetilde{C}(\omega_2,\omega_3,P) |z|^{-2+P^2-(\omega_2+\omega_3)^2+2n}.
\fe
The resulting fully regularized expression for the $1_R\to3_R$ scattering amplitude of type 0B strings is given by
\ie
&\cA_{1_R\to 3_R}(\omega_1,\omega_2,\omega_3)  = i\delta\left(\omega-\sum_{i=1}^3 \omega_i\right)\, g_s^4 C_{S^2} \frac{1}{2} \\
&~\times \Bigg\{ \omega_2^2\omega_3^2\int_\bC d^2z \left[ |z|^{-2\omega_1\omega_2}|1-z|^{-2\omega_2\omega_3-2}\langle V_\omega(\infty) V_{\omega_3}(1) V_{\omega_2}(z,\zbar) V_{\omega_1}(0) \rangle - R_s^{\scriptscriptstyle VVVV} - R_t^{\scriptscriptstyle VVVV} - R_u^{\scriptscriptstyle VVVV} \right]  \\
&~~~~~~ + \int_\bC d^2z \left[ -|z|^{-2\omega_1\omega_2}|1-z|^{-2\omega_2\omega_3}\langle V_\omega(\infty) W_{\omega_3}(1) W_{\omega_2}(z,\zbar) V_{\omega_1}(0) \rangle - R_s^{\scriptscriptstyle VWWV} - R_t^{\scriptscriptstyle VWWV} - R_u^{\scriptscriptstyle VWWV} \right] \\
&~~~~~~ +\omega_2\omega_3 \int_\bC d^2z \left[ -|z|^{-2\omega_1\omega_2}|1-z|^{-2\omega_2\omega_3}\left[\frac{1}{1-\zbar}\langle V_\omega(\infty) \Lambda_{\omega_3}(1) \Lambda_{\omega_2}(z,\zbar) V_{\omega_1}(0) \rangle \right. \right. \\
& ~~~~~~~~~~~~~~~~~~~~~~~~~~~~~~~~~~~~ +\left.\left. \frac{1}{1-z}\langle V_\omega(\infty) \widetilde{\Lambda}_{\omega_3}(1) \widetilde{\Lambda}_{\omega_2}(z,\zbar) V_{\omega_1}(0) \rangle \right] - R_s^{\scriptscriptstyle V\Lambda\Lambda V} - R_t^{\scriptscriptstyle V\Lambda\Lambda V} - R_u^{\scriptscriptstyle V\Lambda\Lambda V} \right] \Bigg\}.
\label{eq:1to3ampreg}
\fe

\bibliographystyle{JHEP}
\bibliography{0B_refs}

\end{document}